\documentclass[prx]{revtex4-2}
\usepackage[x11names, rgb, html]{xcolor}

\definecolor{red}{HTML}{f54b1a}
\definecolor{pink}{HTML}{d19eb1}
\definecolor{orange}{HTML}{d3772e}
\definecolor{yellow}{HTML}{ebe85d}
\definecolor{green}{HTML}{0f6852}
\definecolor{lightblue}{HTML}{01abe9}
\definecolor{darkblue}{HTML}{1b346c}
\definecolor{tan}{HTML}{e5c39e}
\definecolor{darktan}{HTML}{af9e73}
\definecolor{grey}{HTML}{c3ced0}
\definecolor{darkgrey}{HTML}{9dadc4}
\definecolor{black}{HTML}{110d1b}
\definecolor{white}{HTML}{f1f8f1}

\usepackage{hyperref}
\hypersetup{
  colorlinks = true,
  linkcolor = red,
  citecolor = darkblue,
  urlcolor = lightblue
}

\usepackage[cmintegrals,cmbraces]{newtxmath}
\usepackage{graphicx}

\usepackage[margin=1.25in]{geometry}

\usepackage{algorithm}
\usepackage{algpseudocode}
\algrenewcommand{\algorithmiccomment}[1]{$\vartriangleright$ #1}
\algrenewcommand{\algorithmicreturn}{\textbf{Return: }}
\algnewcommand\algorithmicinput{\textbf{Input: }}
\algnewcommand\Input{\State \algorithmicinput}
\algnewcommand\algorithmicforeach{\textbf{for each}}
\algdef{S}[FOR]{ForEach}[1]{\algorithmicforeach\ #1\ \algorithmicdo}

\def\thetab{\boldsymbol{\theta}}

\def\ab{\boldsymbol{a}}

\def\eb{\boldsymbol{e}}

\def\xb{\boldsymbol{x}}
\def\yb{\boldsymbol{y}}

\def\Mb{\boldsymbol{M}}
\def\Tb{\boldsymbol{T}}
\def\Xb{\boldsymbol{X}}

\def\Wb{\boldsymbol{W}}

\def\RR{\mathbb{R}} \def\NN{\mathbb{N}} 
\def\EE{\mathbb{E}}

\def\<{\langle} \def\>{\rangle}

\DeclareRobustCommand{\argmin}{\operatorname*{argmin}}

\begin{document}

\title{Probing the Theoretical and Computational Limits of Dissipative Design}
\author{Shriram Chennakesavalu}
\author{Grant M. Rotskoff}
\email{rotskoff@stanford.edu}
\affiliation{Department of Chemistry, Stanford University}

\begin{abstract}
    Self-assembly, the process by which interacting components form well-defined and often intricate structures, is typically thought of as a spontaneous process arising from equilibrium dynamics.
    When a system is driven by external \emph{nonequilibrium} forces, states statistically inaccessible to the equilibrium dynamics can arise, a process sometimes termed direct self-assembly.
    However, if we fix a given target state and a set of external control variables, it is not well-understood i) how to design a protocol to drive the system towards the desired state nor ii) the energetic cost of persistently perturbing the stationary distribution. 
    Here we derive a bound that relates the proximity to the chosen target with the dissipation associated with the external drive, showing that high-dimensional external control can guide systems towards target distribution but with an inevitable entropic cost.
    Remarkably, the bound holds arbitrarily far from equilibrium. 
    Secondly, we investigate the performance of deep reinforcement learning algorithms and provide evidence for the realizability of complex protocols that stabilize otherwise inaccessible states of matter.
\end{abstract}

\maketitle

\section{Introduction}

Designing molecular materials that robustly and autonomously assemble into specific, targeted mesoscale structures remains a central challenge in a variety of fields, from materials science~\cite{yin_colloidal_2005,ma_inverse_2019,gadelrab_inverting_2017,ronellenfitsch_inverse_2019} to biology~\cite{cameron_biogenesis_2013,sigl_programmable_2021,rotskoff_robust_2018}.
The canonical approach to this design problem is to engineer components with specific molecular interactions that stabilize a thermodynamic ground state corresponding to the target~\cite{furst_directed_2013}, an inverse approach pioneered in Refs.~\cite{rechtsman_optimized_2005,rechtsman_2,rechtsman_3}.
Advances in tuneable materials, such as patchy particles~\cite{yi_recent_2013}, DNA coated colloids~\cite{wang_crystallization_2015}, and DNA origami~\cite{ke_threedimensional_2012} have enabled the realization of highly-specific, directional interactions among the constituent elements and subsequent efforts to optimize these interactions~\cite{chen_patchy_18, Ma_magnetic_self_assembly_2020, romano_designing_2020}. 
However, this paradigm is by no means general---in fact, in many instances, self-assembly is driven not by unique and addressable~\cite{hedges_growth_2014} interactions but rather by weak and nonspecific ones.
Here we explore a distinct approach, one based on \emph{nonequilibrium} external control of a dynamically assembling system, as opposed to designing interactions among the components of a system.

Tailoring self-assembly to produce materials with exotic or desirable properties has long been a goal in the molecular sciences~\cite{furst_directed_2013}.
Early computational work on this topic focused on the inverse design of interaction potentials that produced materials with, for example, target radial distribution functions, densities, and band gaps~\cite{rechtsman_optimized_2005}.
Parameterizing flexible interaction potentials that consistently achieve the desired states is nevertheless challenging. 
The widespread adoption of machine learning techniques in scientific computing has led many to revisit this problem using more sophisticated numerical representations of optimizable potential energy functions.
These techniques have led to significant advances, allowing both for more complex interactions~\cite{das_variational_2021} and new computational approaches~\cite{goodrich_designing_2021}, increasing the set of structures accessible to designer self-assembly.

At molecular scales, it is often impossible to alter the nature of an interaction without fundamentally changing the molecular components as well, potentially disrupting biological or chemical function. 
Rather than viewing self-assembly as a thermodynamic process in which the ultimate structure is determined by a minimum of the free energy, we examine the \emph{dynamics} of assembly trajectories~\cite{furst_directed_2013} and ask whether an external agent can perturb the assembling components so as to maximize the yield of a target structure. 
Because the magnitude of fluctuations at the nanoscale is comparable to the size of the system itself, the task of determining an effective external protocol for control resembles a stochastic optimal control problem. 

Just as interaction design is inherently limited by the constituent materials, the precision of control is dictated by the external fields that couple to a given system and the spatio-temporal resolution with which we can reasonably alter these fields. 
Moreover, an external control approach to directed self-assembly presents new computational challenges: stochastic optimal control problems are typically formulated as high-dimensional partial differential equations, which cannot be solved either analytically or numerically for nontrivial systems.
Here, we instead pose the design problem as the optimization of a Markov decision process~\cite{sutton_reinforcement_2018}, which is in turn amenable to deep reinforcement learning algorithms. 
Of course, these complicated high-dimensional problems have also benefited from advances in deep learning, enabling the optimization of very high-dimensional feedback protocols for essentially arbitrary physical systems. 

In this paper, we investigate the theoretical and computational limits of nonequilibrium control in the context of two minimal models of molecular self-assembly.
We establish theoretically a relationship between the dissipative cost of a protocol and the fidelity with which a target structure can be produced, akin to bounds that have been established for nonequilibrium growth processes~\cite{nguyen_design_2016}.
We then explore the capabilities of deep reinforcement learning algorithms to control the quenched cluster size distribution of a system of particles using a feedback thermal annealing protocol as well as the steady-state cluster size distribution of nonequilibrium actively driven colloids.
Taken together, our theoretical and computational results emphasize that high dimensional control can target assembly outcomes with high precision but with an inescapable dissipative cost.

\section{The dissipative cost of high-fidelity control}
\label{sec:bound}

Consider a physical system with coordinates $\xb\in \Omega\subset \mathbb{T}^d$ evolving according to overdamped Langevin equation
\begin{equation}
    d\Xb_t = b(\Xb_t) dt + \sqrt{2D} d\Wb_t
\end{equation}
where $b$ is a nonequilibrium drift, $D = k_{\rm B}T/\mu$ is the diffusion coefficient, and $\Wb_t$ is a Weiner process in $\RR^d$.
We assume that $\Xb_t$ is ergodic so that there exists a unique stationary probability density $\rho_{\rm ss}:\Omega \to \RR.$

Our goal is to develop a feedback-guided, external control protocol $u$ that pushes the steady state distribution towards a specified target. 
At present, we focus on external driving that can be represented as a force, not a noise term, though we consider both regimes in the subsequent numerical experiments. 
We assume that the external control can be implemented as a spatially-varying external force $u$ leading to the controlled SDE
\begin{equation}
    d\Xb_t^u = [b(\Xb_t^u) + u(\Xb_t^u)]dt + \sqrt{2D} d\Wb_t,
    \label{eq:usde}
\end{equation}
which in turn has an associated a steady state density $\rho_u.$
While the most generic design task requires tuning $u$ to coincide with a target steady state distribution $\rho_*$, it is not clear how to specify a target density function for a large interacting particle system, as we typically characterize these systems instead by some low-dimensional observable.
This more limited description requires setting some target average value of a given observable $f:\Omega \to \RR$.
Let us denote the target value of $f$ by $f_*.$
The optimal controller then solves the minimization problem
\begin{equation}
    u_* = \argmin_u |\EE_u f - f_*|
    \label{eq:minprob}
\end{equation}
where $\EE_u$ denotes the expectation over the controlled process~\eqref{eq:usde} and $\EE_* f\equiv f_*$ denotes the target value of $f$, which we view as the expectation over the unknown target distribution (cf. Appendix~\ref{app:w1control} for a detailed discussion). 

In some cases, the chosen observable $f$ might not be informative about the system.
However, while we seek to carry out this minimization for a particular choice of $f$, if we instead allow $f$ to vary and solve the minimax problem to find the controller that minimizes the mean discrepancy over all functions $g$\footnote{The set of functions $g$ satisfies technical assumptions detailed in Appendix~\ref{app:w1control}.}, then the objective is the Kantorovich-Rubenstein dual formulation of the Wasserstein-1 distance~\cite{villani_topics_2003}.
This metric quantifies the distance between the target distribution and the steady state distribution of the controlled process,
\begin{equation}
    \mathcal{W}_1(\rho_u, \rho_*) = \max_g \min_u |\EE_u g - \EE_* g|.
\end{equation}
The Wasserstein distance is an optimal transport distance on probability distributions, measuring the cost to reallocate mass from one distribution to another.
Applications of optimal transport distances have become widespread in data analysis and machine learning. See \citet{peyre_computational_2018} for an applied perspective. For the fixed observable of interest, the mean discrepancy is bounded above by the Wasserstein distance, which in turn, is bounded by the Kullback-Liebler divergence or relative entropy
\begin{equation}
\begin{aligned}
    \min_u |\EE_u f - f_*| &\leq \mathcal{W}_1(\rho_u, \rho_*), \\
     &\leq C \sqrt{2 D_{\rm KL}(\rho_u, \rho_*)}. 
     \label{eq:klineq}
\end{aligned}
\end{equation}
The first inequality follows from an application of dual formulation of the total variation distance and subsequently an application of Pinsker's inequality.
Interestingly, this upper bound has a direct physical interpretation in terms of the dissipative cost of controlling the trajectory to alter the steady state distribution. 

The Kullback-Leibler divergence between the stationary distribution of the controlled process and the target distribution can be interpreted as a nonequilibrium free energy difference~\cite{sivak_nearequilibrium_2012} between the two distributions. 
Maintaining a nonequilibrium steady state distinct from the steady state of the unguided system incurs a time-extensive entropic cost. 
For the path measures associated with the dynamics, we can express this quantity explicitly via the Girsanov theorem, as shown in Appendix~\ref{app:w1control}.
In particular, we see that the entropic cost of control per unit time can be written as a time-average, quantifying the cost of designing a dissipative steady state.
This design cost can be measured by computing $D_{\rm KL}(\rho_u \| \rho_{\rm ss}).$
For the protocol $u$, we show that the cost of reducing the Kullback-Liebler divergence in \eqref{eq:klineq} to minimize \eqref{eq:minprob} is
\begin{equation}
    \sigma_{\rm ex} = \frac1T \int_0^T u(\Xb_t^u)^2 dt \geq 0;
\end{equation}
that is, there is an inverse relationship between the fidelity of control and the non-negative cost of control.
The optimal rate is given using $u_* = b_* - b$, which is in general unknown and which vanishes when no control is needed.
We can still examine the relationship indirectly by monitoring the total entropy production of the controlled system. 
This relation holds when the original, controlled, and target dynamics are all arbitrarily far from equilibrium.

The relation~\eqref{eq:klineq} helps quantify the dissipative cost of controlling a system, even when the state of the system is only partially observed (i.e., through an observable).
Importantly, the relation that we derive is only an upper bound---the discrepancy could be small for some uninformative choice of $f$ without any significant perturbation.
Similarly, many dissipative processes will not necessarily reduce the discrepancy between the mean and its target value---that is, the control could be imprecise. 
Nevertheless, when this bound is tight, the KL divergence quantifies the physical cost of driving the system towards the externally specified target.

\section{Reinforcement Learning for High-Dimensional Control Protocols}

\begin{figure}[ht]
    \centering
\includegraphics[width=\linewidth]{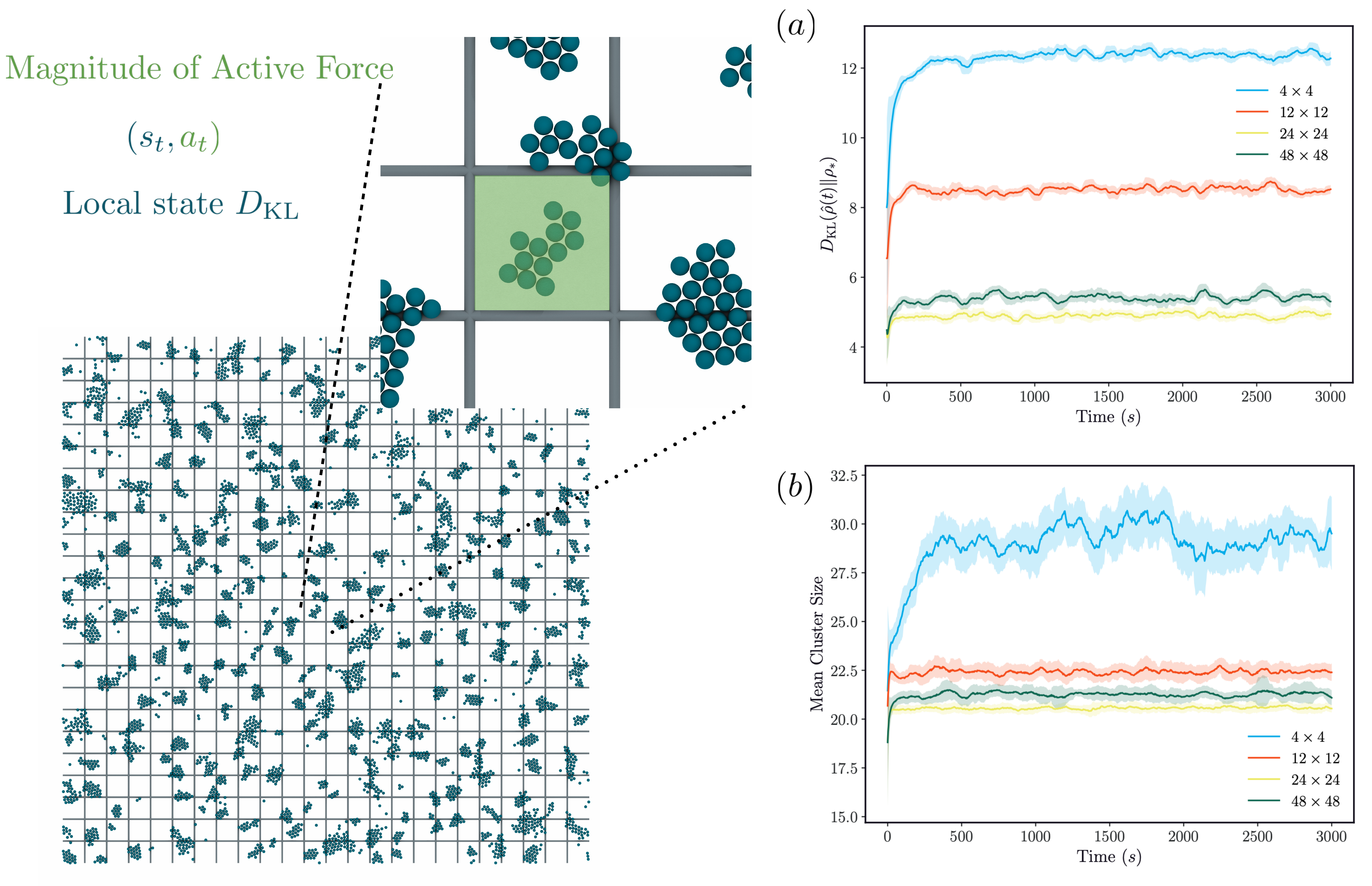}
    \caption{Activity based control of transient motility induced phase separated clusters. Left: the state-action pair of the Markov decision process is depicted graphically. The activity is modulated externally with spatial resolution indicated by the grid. The state is mean cluster size within a region. $(a)$ The KL divergence from instantaneous cluster size distribution to the target distribution as a function of time for optimized protocols with increasing control resolution: $4\times 4$, $12\times 12$, $24\times 24$, and $48\times 48$. $(b)$ The mean value (target $20$) plotted as a function of time for the optimized protocols. }
    \label{fig:active}
\end{figure}

\emph{A priori}, finding the optimal $u$ requires both detailed information about the dynamics of the system and the ability to implement complex interactions.
Fortunately, the minimization problem~\eqref{eq:minprob} is also amenable to model-free reinforcement learning, an approach we explore here.
Moreover, the infimum over protocols $u:\RR^d \to \RR^d$ requires minimizing with respect to arbitrary many-body force functions which are unlikely to be realizable in experimental settings. 
We pursue a more pragmatic approach by building the constraints of control directly into our protocol $u$; these constraints are imposed by restricting to a fixed class of experimentally accessible protocols.

We drive the system so that the observable of interest, a cluster size distribution throughout this paper, matches an externally specified target.
We defined the observable by first introducing a map $h:\RR^d \to \NN^n$ which counts the number of clusters of size $k$ and stores that number in the $k$th entry of an $n$-dimensional vector where $n$ is the total number of particles in the system and hence the maximum cluster size.
We then define the normalized histogram of cluster sizes for the configuration $\hat\rho(h(\xb))$ and measure the discrepancy between this histogram and the target using the Kullback-Leibler divergence, 
\begin{equation}
    \mathcal{C}(\xb) = D_{\rm KL}(\hat \rho(h(\xb)) \| \rho_*).
    \label{eq:reward}
 \end{equation}
 This empirical distribution implicitly depends on the external controller $u$ through the sampled state $\xb$. 
 Importantly, this cost functional makes evaluation of the loss function independent of the particle dynamics, which in turn allows for model free reinforcement learning.
Using a cluster size distribution rather than the average cluster size ensures that the observable robustly describes the target state even when the region of control is large. 
 
Because the objective~\eqref{eq:reward} does not depend explicitly on an unspecified path measure, it is a tractable target for optimization. 
We consider control functions $u$, depicted schematically in Fig.~\ref{fig:active}, in which the external control drives a system locally with a fixed spatial patterning. 
While the protocol is not an arbitrary many-body force, in our case it remains high-dimensional and $u$ may be a complicated function of the configuration $\xb$.
Despite this complexity, neural networks offer a robust, high-dimensional function representation, which we exploit in our representation of $u.$ 
The steady state $\rho_u$ distribution depends on the dynamics of the system, so direct, gradient-based optimization of~\eqref{eq:reward} is challenging. 
In our setup, the duration of the period between protocol updates is sufficiently long that the gradients of the control parameters become too small to meaningfully optimize the objective by backpropagating through the dynamics.
There is also a conceptual reason for employing a model-free optimization algorithm---in experimental settings, we cannot require precise knowledge of the microscopic dynamics of the system to design an external protocol.  

Because the optimal $u$ is time-independent by construction, we use a time-local representation of the joint dynamics of the system and the controller, also known as a Markov decision process.
Optimization problems of this type are the basic framework for reinforcement learning and have been studied extensively in the machine learning and control literature~\cite{sutton_reinforcement_2018}. 
Within this framework, maintaining the optimal steady state distribution requires incorporating information about the expected future divergence from the target distribution, as measured by $\mathcal{C}$.
For a fixed protocol $u$, the cumulative expected future divergence from the target distribution is an expectation over the dynamics of the system, starting from a given state $\xb_t$, is
\begin{equation}
    \bar{\mathcal{C}}(\xb_t, u) = \EE_u^{\xb_t} \sum_{k=0}^\infty \gamma^k \mathcal{C}(\xb_{t+\tau(k+1)})
    \label{eq:value}
\end{equation}
where $\gamma < 1$ is a so-called discount factor that ensures that the sum is convergent and gives additional weight to temporally proximate states. 
Some reinforcement learning algorithms, e.g. policy gradient~\cite{sutton_reinforcement_2018}, use~\eqref{eq:value} as a direct target for optimization, but the lack of time locality makes gradient based optimization challenging when the dynamics occurs over long time scales~\cite{recht_tour_2019}.

Deep reinforcement learning algorithms based on $Q$-learning lift the expected cost $\bar{C}$ so that it depends on a given state-action pair $(\xb_t,\ab_t)$ and protocol $u$.
We assume that the system evolves with $u(\xb_s) = \ab_t$ constant for a fixed duration $s\in[t,t+\tau]$ so that $\xb_{t+\tau}$ is obtained by solving~\eqref{eq:usde} with the initial condition $\xb_t$; this means that the feedback protocol has a time lag and in practice we choose $\tau$ to be sufficiently short that changes in the cost were minimal. 
The lifted cost functional, conventionally called $Q$, quantifies the future cost assuming that the action $\ab_t$ is taken at time $t$, explicitly,
\begin{equation}
\label{eq:Q}
    Q^u(\xb_t, \ab_t) = \EE^{\xb_t}_u \left[ \mathcal{C}(\xb_{t+\tau}; \ab_t) + \sum_{k=1}^\infty \gamma^k \mathcal{C}(\xb_{t+\tau(k+1)}) \right].
\end{equation}
In~\eqref{eq:Q}, the expectation is carried out over a trajectory initialized at $\xb_t$ and subject to the action $\ab_t$ until time $\tau$ and subsequently using the protocol $u.$
The optimal next action for a given protocol $u$ is then simply $\argmin_{\ab} Q^u(\xb_t, \ab).$ 
Employing $Q$-learning requires estimating $Q$ usually with a value iteration algorithm~\cite{watkins_technical_1992}; traditionally this has been carried out using a tabular representation of $Q$, meaning that every state-action pair must be visited by the dynamics in order to provide an accurate representation. 
Of course, if the state or action space is high-dimensional, this is infeasible. 
Recently, alternative strategies that represent $Q$ as a deep neural network have shown promise in a variety of contexts. 
Importantly, when $Q$ is represented as a neural network, high-dimensional state and action spaces become tractable. 
We employ a variant of deep $Q$-learning \cite{mnih-atari-2013, fujimoto_addressing_2018} that uses two deep neural networks to approximate the $Q$ function, called double $Q$ learning which helps avoid minimization bias in the estimate of the minimizer; we discuss the details of this approach in Appendix~\ref{app:dql}.

\section{Controlling Cluster Sizes in Active Colloids}

To test this reinforcement learning approach to dissipative design, we first considered a model colloidal particle system actively driven by externally controlled light sources. 
Models of self-propelled or active matter evince rich phase behavior and have rapidly become canonical models for pattern formation out of equilibrium.
In these systems, when the P\'{e}clet number is sufficiently large, the mobility depends strongly on the local density, which results in nonequilibrium phase separation~\cite{cates_motilityinduced_2015}. 
This phenomenon, called motility induced phase separation (MIPS), requires energy consumption and is largely independent of the inter-particle interactions~\cite{hagan_structure_2013, palacci_living_2013, cates_motilityinduced_2015}.

Because the activity can be externally modulated with simple controls, for example by selectively illuminating a portion of the system with light of variable intensity, it offers a natural example for control.
Reinforcement learning has shown some success in controlling active systems: recently, \citet{falk_learning_2021} examined enhancing transport properties using low-dimensional external protocols optimized with an actor-critic model.
The externally modulated activity leads to clustered states, but when this activity is turned off, the particles diffuse apart and the clusters disintegrate. 
The limitations of this control will necessarily limit the steady state distributions that can be accessed.
In turn, this means that excess dissipation associated with the feedback protocol may enhance control, but it is possible that some of the energy expended goes to waste.

To explore the limits of an activity-inducing dissipative external control, we sought to maintain a steady state distribution consisting of clusters of particles much smaller than the macroscopic aggregate that forms when there is constant activity.
We specified a target distribution $\rho_*$ of cluster sizes as in \eqref{eq:reward} using three distinct target distributions, all with identical mean and variance.
The distribution of cluster sizes is discrete, so we first tested a binomial distribution with $n=20, p=3/4$.
Because this distribution decay rapidly in the tails, we also tested a Gamma distribution $\Gamma(k,\theta)$ with shape parameter $k=25$ and scale parameter $\theta=3/5$ and computed the corresponding probability mass function by integrating the probability density over the bins.
Finally, to remove any effects due to asymmetry of the target distribution, we used a Gaussian target distribution with $\mu=20, \sigma^2=5$.

\begin{figure}
    \centering
    \includegraphics[width=0.75\linewidth]{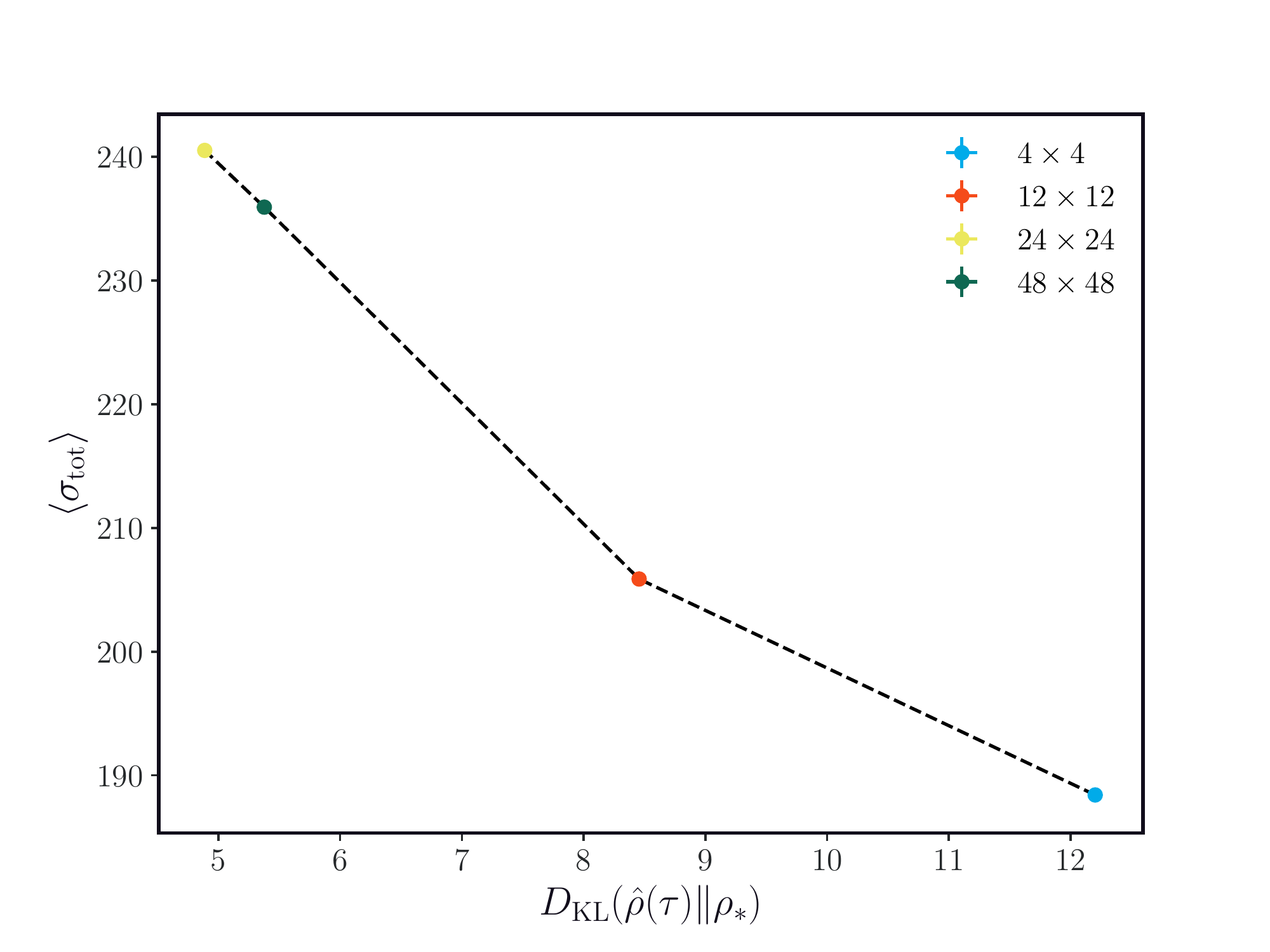}
    \caption{The total entropy production of the controlled system plotted against the fidelity of control as measured by the KL divergence from the target.}
    \label{fig:dissipation}
\end{figure}

We used deep $Q$-learning to optimize an external protocol which controlled the intensity of the activity over a spatial grid, as depicted in Fig.~\ref{fig:active}.
Both the cost function and the corresponding action were evaluated locally; that is, the cluster size distribution was computed over a given region and then the activity was chosen to minimize the estimated $Q$ function given the observed state.
We tested this approach with grids of increasing resolution, corresponding to increasingly fine-tuned spatial control. 
The optimization was carried out for at least 20 ``episodes'' of 350 decisions, until the relative entropy between the instantaneous distribution and the target had converged.\footnote{It is difficult to directly compare the duration of training because the replay buffer grows more quickly for systems with more regions of control}.
Once the protocol had converged, we computed the relative entropy and the mean cluster size over a collection of long test trajectories, shown in Fig.~\ref{fig:active} $(a)$ and $(b)$.
When the number of control regions is small (e.g, $4\times 4$) perturbations to the activity are not localized enough to prevent the formation of large clusters. 
As a result, the mean cluster size is substantially larger than the target. 
On the other hand, the benefits of high-resolution control diminish as the number of regions becomes very large ($48\times 48$).
In effect, once a single region can accommodate only a few clusters of the target size, protocols of increasing resolution achieve the same outcome.
Furthermore, these high-dimensional protocols come with an additional computational or practical burden.

Protocols with sufficient spatial resolution to execute local control perform well (Fig.~\ref{fig:active}), leading to mean values for the average cluster size that are close to the target.
The variance of the empirical distribution of the steady state under the optimized protocol is larger than the target, emphasizing that our relatively coarse external control is still limited.
The timescale over which a cluster diffusively disintegrates in the absence of activity exceeds the time required to form a cluster.
Because the decision period has a fixed duration and activity will favor the formation of large clusters, there is a relative abundance of clusters with a size that exceeds the mean. 
These clusters also contribute to variance in the left tail of the distribution, because dissolving them creates a large number of free particles. 
Nevertheless, as shown in the schematic of Fig.~\ref{fig:active}, the typical clusters are in line with the target distribution and the formation of macroscopic clusters is always avoided.

We examine the generic relationship between dissipative cost and fidelity of control by computing the total entropy production for increasing resolution of control.
In general we found that more control regions led to better control as measured by the cost function~\eqref{eq:reward}.
Fig.~\ref{fig:dissipation} plots the total dissipation $\sigma_{\rm tot}$ as a function of the KL-divergence from the target; each point is averaged over 100 realizations of the optimized feedback protocol.
As the number of control regions increases from $4\times 4$ up to $24\times 24$, the fidelity increases but with a clear increase in dissipation, providing evidence of the utility of the bound~\eqref{eq:klineq}.
At the highest resolutions, diminishing returns become evident because the typical cluster size become comparable to the region itself---in this regime the cost function is essentially exactly the mean discrepancy in~\eqref{eq:minprob}.
For other target cluster size distributions (see Appendix~\ref{app:ac}), we observed a similar trend.
We emphasize that the generic trade-off between dissipation and fidelity is not necessarily a monotonic relationship, as observed for different cost functions in Appendix~\ref{app:ac} because not all dissipative dynamics will lead to improvement of the cost.

\section{Feedback Guided Annealing}

 \begin{figure}
    \begin{center}
    \includegraphics[width=\linewidth]{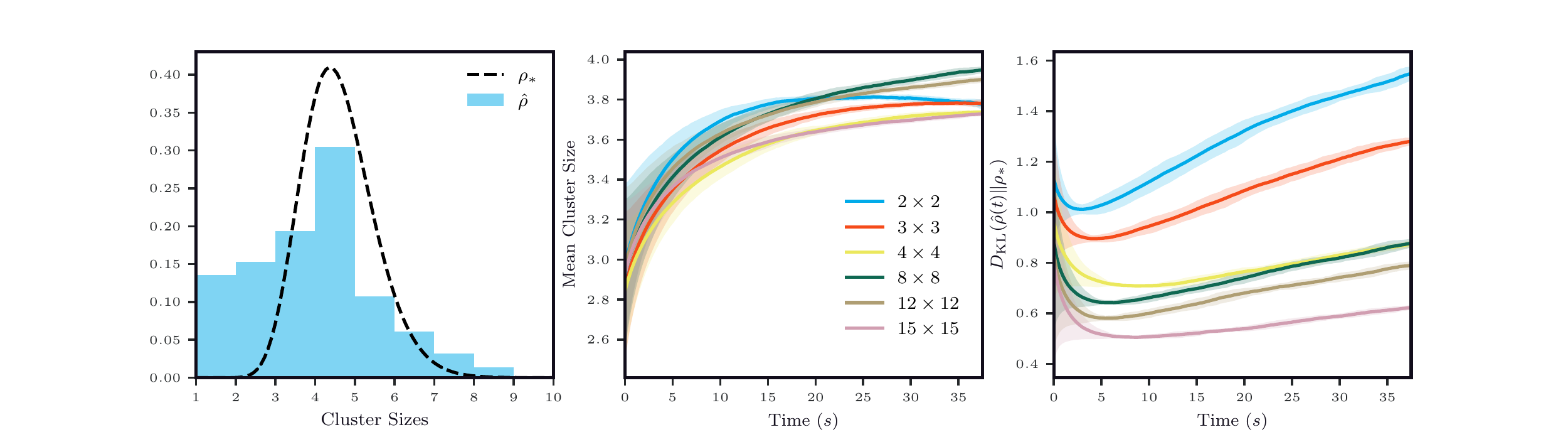}
    \end{center}
     \caption{$(a)$ The target distribution of cluster sizes $\rho_*$ and the empirical distribution obtained using the optimized annealing protocol denoted $\hat \rho.$ While there is an over-representation of isolated particles, as in the case of the active particle system, the model of the distribution and the tail are well-matched by using the protocol. $(b)$ The evolution of the mean cluster size during the feedback annealing process for increasing resolution of temperature control. $(c)$ The evolution of the discrepancy between the instantaneous distribution and the target as a function of time for increasing resolution of temperature control. }
     \label{fig:anneal}
 \end{figure}

While thermal annealing has a long history for macroscopic systems, repeated annealing cycles is an important part of the preparation of a wide variety of nanoscopic materials, from thin films~\cite{makrides_temperature_2012} to DNA origami~\cite{dey_dna_2021}.
Annealing is limited as a mechanism for control because there is essentially only one parameter that can be tuned, the rate at which the temperature is decreased.
This process will eventually find a global free energy minimum, meaning that the ultimate structure is determined entirely by the thermodynamic properties of the system. 
Indeed, there has been significant focus on designing interaction potentials that lead to specific structural motifs in a variety of contexts~\cite{rechtsman_optimized_2005, goodrich_designing_2021, das_variational_2021}.

We examine an alternative paradigm that exercises more localized control with measurement-guided feedback to design an annealing schedule.
Rather than globally tuning a temperature, we locally update the temperature on a grid, see Fig.~\ref{fig:anneal}.
The specific temperature that the protocol prescribes depends on the local configuration.
While the annealing process itself is substantially more complicated in this framework, the approach could be used with arbitrary materials and does not require the realization of highly specific interactions, which can be enormously challenging to engineer in nanoscale systems. 

To assess the prospects of this feedback guided annealing procedure, we studied a minimal example of cluster formation using a 2D Lennard-Jones system. 
At low temperatures, the free energy minimum corresponds to a single cluster, but at low densities, an instantaneous temperature quench from a high-temperature state will yield a kinetically trapped configuration that consists of small clusters.

We sought to control the distribution of these intermediate clusters by optimizing a thermal annealing function using reinforcement learning. 
We fixed a target cluster size distribution, chosen to be a Gamma distribution with variance $\sigma^2=1$ and a mean $\mu=4,$ shown as a dashed line in Fig.~\ref{fig:anneal}.
We trained an external controller that used the local cluster size distribution to determine the subsequent temperature within each region of control.
With only coarse control, the annealing reliably produced distributions of cluster sizes with a mean value close to the target (Fig.~\ref{fig:anneal} (c)).
However, high-resolution control was required to yield a distribution close---as measured by the KL-divergence---to the target distribution.
The empirical cluster size distribution, averaged over 1000 annealing trajectories, is shown in Fig.~\ref{fig:anneal} (a) for a $15\times 15$ grid. 
While there is an over-representation of isolated particles by small amount, the distribution is remarkably close to the target.
For fewer regions, the KL-divergence (Fig.~\ref{fig:anneal} (d)) is substantially larger and the distributions (cf. Appendix~\ref{app:lj}) differ markedly from the target.

Because the external control goes beyond the regime that we treat theoretically in Sec.~\ref{sec:bound}. 
While our analysis extends to systems in which the external control is represented as a drift function $u$, this system relies on changes to the diffusion tensor, which requires a substantially different mathematical treatment, a topic we plan to explore in future work.

\section{Conclusion}
Experimental advances enabling high-resolution external control create new opportunities to produce materials with exotic properties. 
In this work, we seeking to address several fundamental questions about the ``realizability'' of high-fidelity control protocols. 
In doing so, we derive a general bound that establishes a trade-off between the fidelity of control and the dissipative cost of implementing it. 

In many applications, the external fields that could be modulated with high-resolution will be \emph{imprecise}.
That is, we will not necessarily be able to tune a field directly conjugate to the observable of interest.
We do not a priori know how to choose observables that robustly approximate the optimal transport distance, which is a significant topic for future work.
Despite the arguably ``coarse'' control we have---at least compared with the case of interaction design---we nevertheless find that the optimization is successful at producing states with well-defined target mean cluster sizes.
That is, our results demonstrate with appropriate objective functions, reinforcement learning algorithms can identify protocols that closely match target structures even without highly specific interactions. 
What is more, though we do not directly design the protocols to show the relation between dissipation and fidelity, we observe a trend consistent with the prediction throughout. 

The approach we have pursued raises many additional questions. 
Foremost among these perhaps is the tightness of the bound~\eqref{eq:klineq} for a given observable, a question we hope to examine in model systems in future work.
Many other reinforcement learning strategies could be deployed on the systems we studied, including approaches based on policy gradient.
Some of these approaches may enable model-free control for more complicated systems, leading to insights about driven self-assembly in complex environments.

\bibliography{references, references_shriram}

\begin{thebibliography}{39}%
\makeatletter
\providecommand \@ifxundefined [1]{%
 \@ifx{#1\undefined}
}%
\providecommand \@ifnum [1]{%
 \ifnum #1\expandafter \@firstoftwo
 \else \expandafter \@secondoftwo
 \fi
}%
\providecommand \@ifx [1]{%
 \ifx #1\expandafter \@firstoftwo
 \else \expandafter \@secondoftwo
 \fi
}%
\providecommand \natexlab [1]{#1}%
\providecommand \enquote  [1]{``#1''}%
\providecommand \bibnamefont  [1]{#1}%
\providecommand \bibfnamefont [1]{#1}%
\providecommand \citenamefont [1]{#1}%
\providecommand \href@noop [0]{\@secondoftwo}%
\providecommand \href [0]{\begingroup \@sanitize@url \@href}%
\providecommand \@href[1]{\@@startlink{#1}\@@href}%
\providecommand \@@href[1]{\endgroup#1\@@endlink}%
\providecommand \@sanitize@url [0]{\catcode `\\12\catcode `\$12\catcode
  `\&12\catcode `\#12\catcode `\^12\catcode `\_12\catcode `\%12\relax}%
\providecommand \@@startlink[1]{}%
\providecommand \@@endlink[0]{}%
\providecommand \url  [0]{\begingroup\@sanitize@url \@url }%
\providecommand \@url [1]{\endgroup\@href {#1}{\urlprefix }}%
\providecommand \urlprefix  [0]{URL }%
\providecommand \Eprint [0]{\href }%
\providecommand \doibase [0]{https://doi.org/}%
\providecommand \selectlanguage [0]{\@gobble}%
\providecommand \bibinfo  [0]{\@secondoftwo}%
\providecommand \bibfield  [0]{\@secondoftwo}%
\providecommand \translation [1]{[#1]}%
\providecommand \BibitemOpen [0]{}%
\providecommand \bibitemStop [0]{}%
\providecommand \bibitemNoStop [0]{.\EOS\space}%
\providecommand \EOS [0]{\spacefactor3000\relax}%
\providecommand \BibitemShut  [1]{\csname bibitem#1\endcsname}%
\let\auto@bib@innerbib\@empty
%</preamble>
\bibitem [{\citenamefont {Yin}\ and\ \citenamefont
  {Alivisatos}(2005)}]{yin_colloidal_2005}%
  \BibitemOpen
  \bibfield  {author} {\bibinfo {author} {\bibfnamefont {Y.}~\bibnamefont
  {Yin}}\ and\ \bibinfo {author} {\bibfnamefont {A.~P.}\ \bibnamefont
  {Alivisatos}},\ }\href {https://doi.org/10.1038/nature04165} {\bibfield
  {journal} {\bibinfo  {journal} {Nature}\ }\textbf {\bibinfo {volume} {437}},\
  \bibinfo {pages} {664} (\bibinfo {year} {2005})}\BibitemShut {NoStop}%
\bibitem [{\citenamefont {Ma}\ and\ \citenamefont
  {Ferguson}(2019)}]{ma_inverse_2019}%
  \BibitemOpen
  \bibfield  {author} {\bibinfo {author} {\bibfnamefont {Y.}~\bibnamefont
  {Ma}}\ and\ \bibinfo {author} {\bibfnamefont {A.~L.}\ \bibnamefont
  {Ferguson}},\ }\href {https://doi.org/10.1039/C9SM01500K} {\bibfield
  {journal} {\bibinfo  {journal} {Soft Matter}\ }\textbf {\bibinfo {volume}
  {15}},\ \bibinfo {pages} {8808} (\bibinfo {year} {2019})}\BibitemShut
  {NoStop}%
\bibitem [{\citenamefont {Gadelrab}\ \emph {et~al.}(2017)\citenamefont
  {Gadelrab}, \citenamefont {Hannon}, \citenamefont {Ross},\ and\ \citenamefont
  {{Alexander-Katz}}}]{gadelrab_inverting_2017}%
  \BibitemOpen
  \bibfield  {author} {\bibinfo {author} {\bibfnamefont {K.~R.}\ \bibnamefont
  {Gadelrab}}, \bibinfo {author} {\bibfnamefont {A.~F.}\ \bibnamefont
  {Hannon}}, \bibinfo {author} {\bibfnamefont {C.~A.}\ \bibnamefont {Ross}},\
  and\ \bibinfo {author} {\bibfnamefont {A.}~\bibnamefont {{Alexander-Katz}}},\
  }\href {https://doi.org/10.1039/C7ME00062F} {\bibfield  {journal} {\bibinfo
  {journal} {Molecular Systems Design \& Engineering}\ }\textbf {\bibinfo
  {volume} {2}},\ \bibinfo {pages} {539} (\bibinfo {year} {2017})}\BibitemShut
  {NoStop}%
\bibitem [{\citenamefont {Ronellenfitsch}\ \emph {et~al.}(2019)\citenamefont
  {Ronellenfitsch}, \citenamefont {Stoop}, \citenamefont {Yu}, \citenamefont
  {Forrow},\ and\ \citenamefont {Dunkel}}]{ronellenfitsch_inverse_2019}%
  \BibitemOpen
  \bibfield  {author} {\bibinfo {author} {\bibfnamefont {H.}~\bibnamefont
  {Ronellenfitsch}}, \bibinfo {author} {\bibfnamefont {N.}~\bibnamefont
  {Stoop}}, \bibinfo {author} {\bibfnamefont {J.}~\bibnamefont {Yu}}, \bibinfo
  {author} {\bibfnamefont {A.}~\bibnamefont {Forrow}},\ and\ \bibinfo {author}
  {\bibfnamefont {J.}~\bibnamefont {Dunkel}},\ }\href
  {https://doi.org/10.1103/PhysRevMaterials.3.095201} {\bibfield  {journal}
  {\bibinfo  {journal} {Physical Review Materials}\ }\textbf {\bibinfo {volume}
  {3}},\ \bibinfo {pages} {095201} (\bibinfo {year} {2019})}\BibitemShut
  {NoStop}%
\bibitem [{\citenamefont {Cameron}\ \emph {et~al.}(2013)\citenamefont
  {Cameron}, \citenamefont {Wilson}, \citenamefont {Bernstein},\ and\
  \citenamefont {Kerfeld}}]{cameron_biogenesis_2013}%
  \BibitemOpen
  \bibfield  {author} {\bibinfo {author} {\bibfnamefont {J.~C.}\ \bibnamefont
  {Cameron}}, \bibinfo {author} {\bibfnamefont {S.~C.}\ \bibnamefont {Wilson}},
  \bibinfo {author} {\bibfnamefont {S.~L.}\ \bibnamefont {Bernstein}},\ and\
  \bibinfo {author} {\bibfnamefont {C.~A.}\ \bibnamefont {Kerfeld}},\ }\href
  {https://doi.org/10.1016/j.cell.2013.10.044} {\bibfield  {journal} {\bibinfo
  {journal} {Cell}\ }\textbf {\bibinfo {volume} {155}},\ \bibinfo {pages}
  {1131} (\bibinfo {year} {2013})}\BibitemShut {NoStop}%
\bibitem [{\citenamefont {Sigl}\ \emph {et~al.}(2021)\citenamefont {Sigl},
  \citenamefont {Willner}, \citenamefont {Engelen}, \citenamefont {Kretzmann},
  \citenamefont {Sachenbacher}, \citenamefont {Liedl}, \citenamefont {Kolbe},
  \citenamefont {Wilsch}, \citenamefont {Aghvami}, \citenamefont {Protzer},
  \citenamefont {Hagan}, \citenamefont {Fraden},\ and\ \citenamefont
  {Dietz}}]{sigl_programmable_2021}%
  \BibitemOpen
  \bibfield  {author} {\bibinfo {author} {\bibfnamefont {C.}~\bibnamefont
  {Sigl}}, \bibinfo {author} {\bibfnamefont {E.~M.}\ \bibnamefont {Willner}},
  \bibinfo {author} {\bibfnamefont {W.}~\bibnamefont {Engelen}}, \bibinfo
  {author} {\bibfnamefont {J.~A.}\ \bibnamefont {Kretzmann}}, \bibinfo {author}
  {\bibfnamefont {K.}~\bibnamefont {Sachenbacher}}, \bibinfo {author}
  {\bibfnamefont {A.}~\bibnamefont {Liedl}}, \bibinfo {author} {\bibfnamefont
  {F.}~\bibnamefont {Kolbe}}, \bibinfo {author} {\bibfnamefont
  {F.}~\bibnamefont {Wilsch}}, \bibinfo {author} {\bibfnamefont {S.~A.}\
  \bibnamefont {Aghvami}}, \bibinfo {author} {\bibfnamefont {U.}~\bibnamefont
  {Protzer}}, \bibinfo {author} {\bibfnamefont {M.~F.}\ \bibnamefont {Hagan}},
  \bibinfo {author} {\bibfnamefont {S.}~\bibnamefont {Fraden}},\ and\ \bibinfo
  {author} {\bibfnamefont {H.}~\bibnamefont {Dietz}},\ }\href
  {https://doi.org/10.1038/s41563-021-01020-4} {\bibfield  {journal} {\bibinfo
  {journal} {Nature Materials}\ ,\ \bibinfo {pages} {1}} (\bibinfo {year}
  {2021})}\BibitemShut {NoStop}%
\bibitem [{\citenamefont {Rotskoff}\ and\ \citenamefont
  {Geissler}(2018)}]{rotskoff_robust_2018}%
  \BibitemOpen
  \bibfield  {author} {\bibinfo {author} {\bibfnamefont {G.~M.}\ \bibnamefont
  {Rotskoff}}\ and\ \bibinfo {author} {\bibfnamefont {P.~L.}\ \bibnamefont
  {Geissler}},\ }\href {https://doi.org/10.1073/pnas.1802499115} {\bibfield
  {journal} {\bibinfo  {journal} {Proceedings of the National Academy of
  Sciences}\ }\textbf {\bibinfo {volume} {112}},\ \bibinfo {pages} {201802499}
  (\bibinfo {year} {2018})}\BibitemShut {NoStop}%
\bibitem [{\citenamefont {Furst}(2013)}]{furst_directed_2013}%
  \BibitemOpen
  \bibfield  {author} {\bibinfo {author} {\bibfnamefont {E.~M.}\ \bibnamefont
  {Furst}},\ }\href {https://doi.org/10.1039/C3SM90126B} {\bibfield  {journal}
  {\bibinfo  {journal} {Soft Matter}\ }\textbf {\bibinfo {volume} {9}},\
  \bibinfo {pages} {9039} (\bibinfo {year} {2013})}\BibitemShut {NoStop}%
\bibitem [{\citenamefont {Rechtsman}\ \emph {et~al.}(2005)\citenamefont
  {Rechtsman}, \citenamefont {Stillinger},\ and\ \citenamefont
  {Torquato}}]{rechtsman_optimized_2005}%
  \BibitemOpen
  \bibfield  {author} {\bibinfo {author} {\bibfnamefont {M.~C.}\ \bibnamefont
  {Rechtsman}}, \bibinfo {author} {\bibfnamefont {F.~H.}\ \bibnamefont
  {Stillinger}},\ and\ \bibinfo {author} {\bibfnamefont {S.}~\bibnamefont
  {Torquato}},\ }\href {https://doi.org/10.1103/PhysRevLett.95.228301}
  {\bibfield  {journal} {\bibinfo  {journal} {Physical Review Letters}\
  }\textbf {\bibinfo {volume} {95}},\ \bibinfo {pages} {228301} (\bibinfo
  {year} {2005})}\BibitemShut {NoStop}%
\bibitem [{\citenamefont {Rechtsman}\ \emph
  {et~al.}(2006{\natexlab{a}})\citenamefont {Rechtsman}, \citenamefont
  {Stillinger},\ and\ \citenamefont {Torquato}}]{rechtsman_2}%
  \BibitemOpen
  \bibfield  {author} {\bibinfo {author} {\bibfnamefont {M.}~\bibnamefont
  {Rechtsman}}, \bibinfo {author} {\bibfnamefont {F.}~\bibnamefont
  {Stillinger}},\ and\ \bibinfo {author} {\bibfnamefont {S.}~\bibnamefont
  {Torquato}},\ }\href {https://doi.org/10.1103/PhysRevE.73.011406} {\bibfield
  {journal} {\bibinfo  {journal} {Physical Review E}\ }\textbf {\bibinfo
  {volume} {73}},\ \bibinfo {pages} {011406} (\bibinfo {year}
  {2006}{\natexlab{a}})},\ \bibinfo {note} {number of pages: 12 Publisher:
  American Physical Society}\BibitemShut {NoStop}%
\bibitem [{\citenamefont {Rechtsman}\ \emph
  {et~al.}(2006{\natexlab{b}})\citenamefont {Rechtsman}, \citenamefont
  {Stillinger},\ and\ \citenamefont {Torquato}}]{rechtsman_3}%
  \BibitemOpen
  \bibfield  {author} {\bibinfo {author} {\bibfnamefont {M.~C.}\ \bibnamefont
  {Rechtsman}}, \bibinfo {author} {\bibfnamefont {F.~H.}\ \bibnamefont
  {Stillinger}},\ and\ \bibinfo {author} {\bibfnamefont {S.}~\bibnamefont
  {Torquato}},\ }\href {https://doi.org/10.1103/PhysRevE.74.021404} {\bibfield
  {journal} {\bibinfo  {journal} {Physical Review E}\ }\textbf {\bibinfo
  {volume} {74}},\ \bibinfo {pages} {021404} (\bibinfo {year}
  {2006}{\natexlab{b}})},\ \bibinfo {note} {number of pages: 7 Publisher:
  American Physical Society}\BibitemShut {NoStop}%
\bibitem [{\citenamefont {Yi}\ \emph {et~al.}(2013)\citenamefont {Yi},
  \citenamefont {Pine},\ and\ \citenamefont {Sacanna}}]{yi_recent_2013}%
  \BibitemOpen
  \bibfield  {author} {\bibinfo {author} {\bibfnamefont {G.-R.}\ \bibnamefont
  {Yi}}, \bibinfo {author} {\bibfnamefont {D.~J.}\ \bibnamefont {Pine}},\ and\
  \bibinfo {author} {\bibfnamefont {S.}~\bibnamefont {Sacanna}},\ }\href
  {https://doi.org/10.1088/0953-8984/25/19/193101} {\bibfield  {journal}
  {\bibinfo  {journal} {J. Phys.: Condens. Matter}\ }\textbf {\bibinfo {volume}
  {25}},\ \bibinfo {pages} {193101} (\bibinfo {year} {2013})}\BibitemShut
  {NoStop}%
\bibitem [{\citenamefont {Wang}\ \emph {et~al.}(2015)\citenamefont {Wang},
  \citenamefont {Wang}, \citenamefont {Zheng}, \citenamefont {Ducrot},
  \citenamefont {Yodh}, \citenamefont {Weck},\ and\ \citenamefont
  {Pine}}]{wang_crystallization_2015}%
  \BibitemOpen
  \bibfield  {author} {\bibinfo {author} {\bibfnamefont {Y.}~\bibnamefont
  {Wang}}, \bibinfo {author} {\bibfnamefont {Y.}~\bibnamefont {Wang}}, \bibinfo
  {author} {\bibfnamefont {X.}~\bibnamefont {Zheng}}, \bibinfo {author}
  {\bibfnamefont {{\'E}.}~\bibnamefont {Ducrot}}, \bibinfo {author}
  {\bibfnamefont {J.~S.}\ \bibnamefont {Yodh}}, \bibinfo {author}
  {\bibfnamefont {M.}~\bibnamefont {Weck}},\ and\ \bibinfo {author}
  {\bibfnamefont {D.~J.}\ \bibnamefont {Pine}},\ }\href
  {https://doi.org/10.1038/ncomms8253} {\bibfield  {journal} {\bibinfo
  {journal} {Nat. Commun.}\ }\textbf {\bibinfo {volume} {6}},\ \bibinfo {pages}
  {7253} (\bibinfo {year} {2015})}\BibitemShut {NoStop}%
\bibitem [{\citenamefont {Ke}\ \emph {et~al.}(2012)\citenamefont {Ke},
  \citenamefont {Ong}, \citenamefont {Shih},\ and\ \citenamefont
  {Yin}}]{ke_threedimensional_2012}%
  \BibitemOpen
  \bibfield  {author} {\bibinfo {author} {\bibfnamefont {Y.}~\bibnamefont
  {Ke}}, \bibinfo {author} {\bibfnamefont {L.~L.}\ \bibnamefont {Ong}},
  \bibinfo {author} {\bibfnamefont {W.~M.}\ \bibnamefont {Shih}},\ and\
  \bibinfo {author} {\bibfnamefont {P.}~\bibnamefont {Yin}},\ }\href
  {https://doi.org/10.1126/science.1227268} {\bibfield  {journal} {\bibinfo
  {journal} {Science}\ }\textbf {\bibinfo {volume} {338}},\ \bibinfo {pages}
  {1177} (\bibinfo {year} {2012})}\BibitemShut {NoStop}%
\bibitem [{\citenamefont {Chen}\ \emph {et~al.}(2018)\citenamefont {Chen},
  \citenamefont {Zhang},\ and\ \citenamefont {Torquato}}]{chen_patchy_18}%
  \BibitemOpen
  \bibfield  {author} {\bibinfo {author} {\bibfnamefont {D.}~\bibnamefont
  {Chen}}, \bibinfo {author} {\bibfnamefont {G.}~\bibnamefont {Zhang}},\ and\
  \bibinfo {author} {\bibfnamefont {S.}~\bibnamefont {Torquato}},\ }\href
  {https://doi.org/10.1021/acs.jpcb.8b05627} {\bibfield  {journal} {\bibinfo
  {journal} {The Journal of Physical Chemistry B}\ }\textbf {\bibinfo {volume}
  {122}},\ \bibinfo {pages} {8462} (\bibinfo {year} {2018})},\ \bibinfo {note}
  {tex.eprint: https://doi.org/10.1021/acs.jpcb.8b05627}\BibitemShut {NoStop}%
\bibitem [{\citenamefont {Ma}\ \emph {et~al.}(2020)\citenamefont {Ma},
  \citenamefont {Lomba},\ and\ \citenamefont
  {Torquato}}]{Ma_magnetic_self_assembly_2020}%
  \BibitemOpen
  \bibfield  {author} {\bibinfo {author} {\bibfnamefont {Z.}~\bibnamefont
  {Ma}}, \bibinfo {author} {\bibfnamefont {E.}~\bibnamefont {Lomba}},\ and\
  \bibinfo {author} {\bibfnamefont {S.}~\bibnamefont {Torquato}},\ }\href
  {https://doi.org/10.1103/PhysRevLett.125.068002} {\bibfield  {journal}
  {\bibinfo  {journal} {Physical Review Letters}\ }\textbf {\bibinfo {volume}
  {125}},\ \bibinfo {pages} {068002} (\bibinfo {year} {2020})},\ \bibinfo
  {note} {number of pages: 6 Publisher: American Physical Society}\BibitemShut
  {NoStop}%
\bibitem [{\citenamefont {Romano}\ \emph {et~al.}(2020)\citenamefont {Romano},
  \citenamefont {Russo}, \citenamefont {Kroc},\ and\ \citenamefont {{\v
  S}ulc}}]{romano_designing_2020}%
  \BibitemOpen
  \bibfield  {author} {\bibinfo {author} {\bibfnamefont {F.}~\bibnamefont
  {Romano}}, \bibinfo {author} {\bibfnamefont {J.}~\bibnamefont {Russo}},
  \bibinfo {author} {\bibfnamefont {L.}~\bibnamefont {Kroc}},\ and\ \bibinfo
  {author} {\bibfnamefont {P.}~\bibnamefont {{\v S}ulc}},\ }\href
  {https://doi.org/10.1103/PhysRevLett.125.118003} {\bibfield  {journal}
  {\bibinfo  {journal} {Physical Review Letters}\ }\textbf {\bibinfo {volume}
  {125}},\ \bibinfo {pages} {118003} (\bibinfo {year} {2020})}\BibitemShut
  {NoStop}%
\bibitem [{\citenamefont {Hedges}\ \emph {et~al.}(2014)\citenamefont {Hedges},
  \citenamefont {Mannige},\ and\ \citenamefont
  {Whitelam}}]{hedges_growth_2014}%
  \BibitemOpen
  \bibfield  {author} {\bibinfo {author} {\bibfnamefont {L.~O.}\ \bibnamefont
  {Hedges}}, \bibinfo {author} {\bibfnamefont {R.~V.}\ \bibnamefont
  {Mannige}},\ and\ \bibinfo {author} {\bibfnamefont {S.}~\bibnamefont
  {Whitelam}},\ }\href {https://doi.org/10.1039/C4SM01021C} {\bibfield
  {journal} {\bibinfo  {journal} {Soft Matter}\ }\textbf {\bibinfo {volume}
  {10}},\ \bibinfo {pages} {6404} (\bibinfo {year} {2014})}\BibitemShut
  {NoStop}%
\bibitem [{\citenamefont {Das}\ and\ \citenamefont
  {Limmer}(2021)}]{das_variational_2021}%
  \BibitemOpen
  \bibfield  {author} {\bibinfo {author} {\bibfnamefont {A.}~\bibnamefont
  {Das}}\ and\ \bibinfo {author} {\bibfnamefont {D.~T.}\ \bibnamefont
  {Limmer}},\ }\href {https://doi.org/10.1063/5.0038652} {\bibfield  {journal}
  {\bibinfo  {journal} {The Journal of Chemical Physics}\ }\textbf {\bibinfo
  {volume} {154}},\ \bibinfo {pages} {014107} (\bibinfo {year}
  {2021})}\BibitemShut {NoStop}%
\bibitem [{\citenamefont {Goodrich}\ \emph {et~al.}(2021)\citenamefont
  {Goodrich}, \citenamefont {King}, \citenamefont {Schoenholz}, \citenamefont
  {Cubuk},\ and\ \citenamefont {Brenner}}]{goodrich_designing_2021}%
  \BibitemOpen
  \bibfield  {author} {\bibinfo {author} {\bibfnamefont {C.~P.}\ \bibnamefont
  {Goodrich}}, \bibinfo {author} {\bibfnamefont {E.~M.}\ \bibnamefont {King}},
  \bibinfo {author} {\bibfnamefont {S.~S.}\ \bibnamefont {Schoenholz}},
  \bibinfo {author} {\bibfnamefont {E.~D.}\ \bibnamefont {Cubuk}},\ and\
  \bibinfo {author} {\bibfnamefont {M.~P.}\ \bibnamefont {Brenner}},\ }\href
  {https://doi.org/10.1073/pnas.2024083118} {\bibfield  {journal} {\bibinfo
  {journal} {Proceedings of the National Academy of Sciences}\ }\textbf
  {\bibinfo {volume} {118}},\ \bibinfo {pages} {e2024083118} (\bibinfo {year}
  {2021})}\BibitemShut {NoStop}%
\bibitem [{\citenamefont {Sutton}\ and\ \citenamefont
  {Barto}(2018)}]{sutton_reinforcement_2018}%
  \BibitemOpen
  \bibfield  {author} {\bibinfo {author} {\bibfnamefont {R.~S.}\ \bibnamefont
  {Sutton}}\ and\ \bibinfo {author} {\bibfnamefont {A.~G.}\ \bibnamefont
  {Barto}},\ }\href@noop {} {\emph {\bibinfo {title} {Reinforcement Learning:
  An Introduction}}},\ \bibinfo {edition} {second edition}\ ed.,\ Adaptive
  Computation and Machine Learning Series\ (\bibinfo  {publisher} {{The MIT
  Press}},\ \bibinfo {address} {{Cambridge, Massachusetts}},\ \bibinfo {year}
  {2018})\BibitemShut {NoStop}%
\bibitem [{\citenamefont {Nguyen}\ and\ \citenamefont
  {Vaikuntanathan}(2016)}]{nguyen_design_2016}%
  \BibitemOpen
  \bibfield  {author} {\bibinfo {author} {\bibfnamefont {M.}~\bibnamefont
  {Nguyen}}\ and\ \bibinfo {author} {\bibfnamefont {S.}~\bibnamefont
  {Vaikuntanathan}},\ }\href {https://doi.org/10.1073/pnas.1609983113}
  {\bibfield  {journal} {\bibinfo  {journal} {Proceedings of the National
  Academy of Sciences}\ }\textbf {\bibinfo {volume} {113}},\ \bibinfo {pages}
  {14231} (\bibinfo {year} {2016})}\BibitemShut {NoStop}%
\bibitem [{Note1()}]{Note1}%
  \BibitemOpen
  \bibinfo {note} {The set of functions $g$ satisfies technical assumptions
  detailed in Appendix~\ref {app:w1control}.}\BibitemShut {Stop}%
\bibitem [{\citenamefont {Villani}(2003)}]{villani_topics_2003}%
  \BibitemOpen
  \bibfield  {author} {\bibinfo {author} {\bibfnamefont {C.}~\bibnamefont
  {Villani}},\ }\href@noop {} {\emph {\bibinfo {title} {Topics in {{Optimal
  Transportation}}}}},\ \bibinfo {series} {Graduate {{Studies}} in
  {{Mathematics}}}\ No.~\bibinfo {number} {58}\ (\bibinfo  {publisher}
  {{American Mathematical Society}},\ \bibinfo {address} {{Providence, Rhode
  Island}},\ \bibinfo {year} {2003})\BibitemShut {NoStop}%
\bibitem [{\citenamefont {Peyr{\'e}}\ and\ \citenamefont
  {Cuturi}(2018)}]{peyre_computational_2018}%
  \BibitemOpen
  \bibfield  {author} {\bibinfo {author} {\bibfnamefont {G.}~\bibnamefont
  {Peyr{\'e}}}\ and\ \bibinfo {author} {\bibfnamefont {M.}~\bibnamefont
  {Cuturi}},\ }\href@noop {} {\bibfield  {journal} {\bibinfo  {journal}
  {arXiv:1803.00567 [stat]}\ } (\bibinfo {year} {2018})},\ \Eprint
  {https://arxiv.org/abs/1803.00567} {arXiv:1803.00567 [stat]} \BibitemShut
  {NoStop}%
\bibitem [{\citenamefont {Sivak}\ and\ \citenamefont
  {Crooks}(2012)}]{sivak_nearequilibrium_2012}%
  \BibitemOpen
  \bibfield  {author} {\bibinfo {author} {\bibfnamefont {D.~A.}\ \bibnamefont
  {Sivak}}\ and\ \bibinfo {author} {\bibfnamefont {G.~E.}\ \bibnamefont
  {Crooks}},\ }\href {https://doi.org/10.1103/PhysRevLett.108.150601}
  {\bibfield  {journal} {\bibinfo  {journal} {Physical Review Letters}\
  }\textbf {\bibinfo {volume} {108}},\ \bibinfo {pages} {150601} (\bibinfo
  {year} {2012})}\BibitemShut {NoStop}%
\bibitem [{\citenamefont {Recht}(2019)}]{recht_tour_2019}%
  \BibitemOpen
  \bibfield  {author} {\bibinfo {author} {\bibfnamefont {B.}~\bibnamefont
  {Recht}},\ }\href {https://doi.org/10.1146/annurev-control-053018-023825}
  {\bibfield  {journal} {\bibinfo  {journal} {Annual Review of Control,
  Robotics, and Autonomous Systems}\ }\textbf {\bibinfo {volume} {2}},\
  \bibinfo {pages} {253} (\bibinfo {year} {2019})}\BibitemShut {NoStop}%
\bibitem [{\citenamefont {Watkins}\ and\ \citenamefont
  {Dayan}(1992)}]{watkins_technical_1992}%
  \BibitemOpen
  \bibfield  {author} {\bibinfo {author} {\bibfnamefont {C.~J. C.~H.}\
  \bibnamefont {Watkins}}\ and\ \bibinfo {author} {\bibfnamefont
  {P.}~\bibnamefont {Dayan}},\ }in\ \href {https://doi.org/10.1007/BF00992698}
  {\emph {\bibinfo {booktitle} {Reinforcement Learning}}},\ \bibinfo {editor}
  {edited by\ \bibinfo {editor} {\bibfnamefont {R.~S.}\ \bibnamefont
  {Sutton}}}\ (\bibinfo  {publisher} {{Springer US}},\ \bibinfo {address}
  {{Boston, MA}},\ \bibinfo {year} {1992})\ pp.\ \bibinfo {pages}
  {55--68}\BibitemShut {NoStop}%
\bibitem [{\citenamefont {Mnih}\ \emph {et~al.}(2013)\citenamefont {Mnih},
  \citenamefont {Kavukcuoglu}, \citenamefont {Silver}, \citenamefont {Graves},
  \citenamefont {Antonoglou}, \citenamefont {Wierstra},\ and\ \citenamefont
  {Riedmiller}}]{mnih-atari-2013}%
  \BibitemOpen
  \bibfield  {author} {\bibinfo {author} {\bibfnamefont {V.}~\bibnamefont
  {Mnih}}, \bibinfo {author} {\bibfnamefont {K.}~\bibnamefont {Kavukcuoglu}},
  \bibinfo {author} {\bibfnamefont {D.}~\bibnamefont {Silver}}, \bibinfo
  {author} {\bibfnamefont {A.}~\bibnamefont {Graves}}, \bibinfo {author}
  {\bibfnamefont {I.}~\bibnamefont {Antonoglou}}, \bibinfo {author}
  {\bibfnamefont {D.}~\bibnamefont {Wierstra}},\ and\ \bibinfo {author}
  {\bibfnamefont {M.}~\bibnamefont {Riedmiller}},\ }in\ \href@noop {} {\emph
  {\bibinfo {booktitle} {Deep Learning Workshop: Advances in Neural Information
  Processing Systems}}},\ Vol.~\bibinfo {volume} {26},\ \bibinfo {editor}
  {edited by\ \bibinfo {editor} {\bibfnamefont {C.~J.~C.}\ \bibnamefont
  {Burges}}, \bibinfo {editor} {\bibfnamefont {L.}~\bibnamefont {Bottou}},
  \bibinfo {editor} {\bibfnamefont {M.}~\bibnamefont {Welling}}, \bibinfo
  {editor} {\bibfnamefont {Z.}~\bibnamefont {Ghahramani}},\ and\ \bibinfo
  {editor} {\bibfnamefont {K.~Q.}\ \bibnamefont {Weinberger}}}\ (\bibinfo
  {publisher} {Curran Associates, Inc.},\ \bibinfo {year} {2013})\BibitemShut
  {NoStop}%
\bibitem [{\citenamefont {Fujimoto}\ \emph {et~al.}(2018)\citenamefont
  {Fujimoto}, \citenamefont {{van Hoof}},\ and\ \citenamefont
  {Meger}}]{fujimoto_addressing_2018}%
  \BibitemOpen
  \bibfield  {author} {\bibinfo {author} {\bibfnamefont {S.}~\bibnamefont
  {Fujimoto}}, \bibinfo {author} {\bibfnamefont {H.}~\bibnamefont {{van
  Hoof}}},\ and\ \bibinfo {author} {\bibfnamefont {D.}~\bibnamefont {Meger}},\
  }in\ \href@noop {} {\emph {\bibinfo {booktitle} {Proceedings of the 35th
  International Conference on Machine Learning}}},\ \bibinfo {series}
  {Proceedings of Machine Learning Research}, Vol.~\bibinfo {volume} {80},\
  \bibinfo {editor} {edited by\ \bibinfo {editor} {\bibfnamefont
  {J.}~\bibnamefont {Dy}}\ and\ \bibinfo {editor} {\bibfnamefont
  {A.}~\bibnamefont {Krause}}}\ (\bibinfo  {publisher} {{PMLR}},\ \bibinfo
  {year} {2018})\ pp.\ \bibinfo {pages} {1587--1596}\BibitemShut {NoStop}%
\bibitem [{\citenamefont {Cates}\ and\ \citenamefont
  {Tailleur}(2015)}]{cates_motilityinduced_2015}%
  \BibitemOpen
  \bibfield  {author} {\bibinfo {author} {\bibfnamefont {M.~E.}\ \bibnamefont
  {Cates}}\ and\ \bibinfo {author} {\bibfnamefont {J.}~\bibnamefont
  {Tailleur}},\ }\href
  {https://doi.org/10.1146/annurev-conmatphys-031214-014710} {\bibfield
  {journal} {\bibinfo  {journal} {Annual Review of Condensed Matter Physics}\
  }\textbf {\bibinfo {volume} {6}},\ \bibinfo {pages} {219} (\bibinfo {year}
  {2015})}\BibitemShut {NoStop}%
\bibitem [{\citenamefont {Hagan}\ \emph {et~al.}(2013)\citenamefont {Hagan},
  \citenamefont {Baskaran},\ and\ \citenamefont
  {Redner}}]{hagan_structure_2013}%
  \BibitemOpen
  \bibfield  {author} {\bibinfo {author} {\bibfnamefont {M.~F.}\ \bibnamefont
  {Hagan}}, \bibinfo {author} {\bibfnamefont {A.}~\bibnamefont {Baskaran}},\
  and\ \bibinfo {author} {\bibfnamefont {G.~S.}\ \bibnamefont {Redner}},\
  }\href {https://doi.org/10.1103/PhysRevLett.110.055701} {\bibfield  {journal}
  {\bibinfo  {journal} {Physical Review Letters}\ }\textbf {\bibinfo {volume}
  {110}},\ \bibinfo {pages} {055701} (\bibinfo {year} {2013})}\BibitemShut
  {NoStop}%
\bibitem [{\citenamefont {Palacci}\ \emph {et~al.}(2013)\citenamefont
  {Palacci}, \citenamefont {Sacanna}, \citenamefont {Steinberg}, \citenamefont
  {Pine},\ and\ \citenamefont {Chaikin}}]{palacci_living_2013}%
  \BibitemOpen
  \bibfield  {author} {\bibinfo {author} {\bibfnamefont {J.}~\bibnamefont
  {Palacci}}, \bibinfo {author} {\bibfnamefont {S.}~\bibnamefont {Sacanna}},
  \bibinfo {author} {\bibfnamefont {A.~P.}\ \bibnamefont {Steinberg}}, \bibinfo
  {author} {\bibfnamefont {D.~J.}\ \bibnamefont {Pine}},\ and\ \bibinfo
  {author} {\bibfnamefont {P.~M.}\ \bibnamefont {Chaikin}},\ }\href
  {https://doi.org/10.1126/science.1230020} {\bibfield  {journal} {\bibinfo
  {journal} {Science}\ }\textbf {\bibinfo {volume} {339}},\ \bibinfo {pages}
  {936} (\bibinfo {year} {2013})}\BibitemShut {NoStop}%
\bibitem [{\citenamefont {Falk}\ \emph {et~al.}(2021)\citenamefont {Falk},
  \citenamefont {Alizadehyazdi}, \citenamefont {Jaeger},\ and\ \citenamefont
  {Murugan}}]{falk_learning_2021}%
  \BibitemOpen
  \bibfield  {author} {\bibinfo {author} {\bibfnamefont {M.~J.}\ \bibnamefont
  {Falk}}, \bibinfo {author} {\bibfnamefont {V.}~\bibnamefont {Alizadehyazdi}},
  \bibinfo {author} {\bibfnamefont {H.}~\bibnamefont {Jaeger}},\ and\ \bibinfo
  {author} {\bibfnamefont {A.}~\bibnamefont {Murugan}},\ }\href@noop {}
  {\bibfield  {journal} {\bibinfo  {journal} {arXiv:2105.04641 [cond-mat]}\ }
  (\bibinfo {year} {2021})},\ \Eprint {https://arxiv.org/abs/2105.04641}
  {arXiv:2105.04641 [cond-mat]} \BibitemShut {NoStop}%
\bibitem [{Note2()}]{Note2}%
  \BibitemOpen
  \bibinfo {note} {It is difficult to directly compare the duration of training
  because the replay buffer grows more quickly for systems with more regions of
  control}\BibitemShut {NoStop}%
\bibitem [{\citenamefont {Makrides}\ \emph {et~al.}(2012)\citenamefont
  {Makrides}, \citenamefont {Zinsser}, \citenamefont {Phinikarides},
  \citenamefont {Schubert},\ and\ \citenamefont
  {Georghiou}}]{makrides_temperature_2012}%
  \BibitemOpen
  \bibfield  {author} {\bibinfo {author} {\bibfnamefont {G.}~\bibnamefont
  {Makrides}}, \bibinfo {author} {\bibfnamefont {B.}~\bibnamefont {Zinsser}},
  \bibinfo {author} {\bibfnamefont {A.}~\bibnamefont {Phinikarides}}, \bibinfo
  {author} {\bibfnamefont {M.}~\bibnamefont {Schubert}},\ and\ \bibinfo
  {author} {\bibfnamefont {G.~E.}\ \bibnamefont {Georghiou}},\ }\href
  {https://doi.org/10.1016/j.renene.2011.11.046} {\bibfield  {journal}
  {\bibinfo  {journal} {Renewable Energy}\ }\textbf {\bibinfo {volume} {43}},\
  \bibinfo {pages} {407} (\bibinfo {year} {2012})}\BibitemShut {NoStop}%
\bibitem [{\citenamefont {Dey}\ \emph {et~al.}(2021)\citenamefont {Dey},
  \citenamefont {Fan}, \citenamefont {Gothelf}, \citenamefont {Li},
  \citenamefont {Lin}, \citenamefont {Liu}, \citenamefont {Liu}, \citenamefont
  {Nijenhuis}, \citenamefont {Sacc{\`a}}, \citenamefont {Simmel}, \citenamefont
  {Yan},\ and\ \citenamefont {Zhan}}]{dey_dna_2021}%
  \BibitemOpen
  \bibfield  {author} {\bibinfo {author} {\bibfnamefont {S.}~\bibnamefont
  {Dey}}, \bibinfo {author} {\bibfnamefont {C.}~\bibnamefont {Fan}}, \bibinfo
  {author} {\bibfnamefont {K.~V.}\ \bibnamefont {Gothelf}}, \bibinfo {author}
  {\bibfnamefont {J.}~\bibnamefont {Li}}, \bibinfo {author} {\bibfnamefont
  {C.}~\bibnamefont {Lin}}, \bibinfo {author} {\bibfnamefont {L.}~\bibnamefont
  {Liu}}, \bibinfo {author} {\bibfnamefont {N.}~\bibnamefont {Liu}}, \bibinfo
  {author} {\bibfnamefont {M.~A.~D.}\ \bibnamefont {Nijenhuis}}, \bibinfo
  {author} {\bibfnamefont {B.}~\bibnamefont {Sacc{\`a}}}, \bibinfo {author}
  {\bibfnamefont {F.~C.}\ \bibnamefont {Simmel}}, \bibinfo {author}
  {\bibfnamefont {H.}~\bibnamefont {Yan}},\ and\ \bibinfo {author}
  {\bibfnamefont {P.}~\bibnamefont {Zhan}},\ }\href
  {https://doi.org/10.1038/s43586-020-00009-8} {\bibfield  {journal} {\bibinfo
  {journal} {Nature Reviews Methods Primers}\ }\textbf {\bibinfo {volume}
  {1}},\ \bibinfo {pages} {1} (\bibinfo {year} {2021})}\BibitemShut {NoStop}%
\bibitem [{\citenamefont {Gibbs}\ and\ \citenamefont
  {Su}(2002)}]{gibbs_choosing_2002}%
  \BibitemOpen
  \bibfield  {author} {\bibinfo {author} {\bibfnamefont {A.~L.}\ \bibnamefont
  {Gibbs}}\ and\ \bibinfo {author} {\bibfnamefont {F.~E.}\ \bibnamefont {Su}},\
  }\href {https://doi.org/10.1111/j.1751-5823.2002.tb00178.x} {\bibfield
  {journal} {\bibinfo  {journal} {International Statistical Review}\ }\textbf
  {\bibinfo {volume} {70}},\ \bibinfo {pages} {419} (\bibinfo {year}
  {2002})}\BibitemShut {NoStop}%
\bibitem [{\citenamefont {Pantazis}\ and\ \citenamefont
  {Katsoulakis}(2013)}]{pantazis_relative_2013}%
  \BibitemOpen
  \bibfield  {author} {\bibinfo {author} {\bibfnamefont {Y.}~\bibnamefont
  {Pantazis}}\ and\ \bibinfo {author} {\bibfnamefont {M.~A.}\ \bibnamefont
  {Katsoulakis}},\ }\href {https://doi.org/10.1063/1.4789612} {\bibfield
  {journal} {\bibinfo  {journal} {The Journal of Chemical Physics}\ }\textbf
  {\bibinfo {volume} {138}},\ \bibinfo {pages} {054115} (\bibinfo {year}
  {2013})}\BibitemShut {NoStop}%
\end{thebibliography}%

\appendix

\newpage
\section{Relation between coarse-grained control and dissipation}\label{app:w1control}

Let $\xb \in \Omega \subset \mathbb{T}^d$ denote a configuration on the system, taking coordinates on the torus due to periodic boundary conditions.
We set
\begin{equation}
    C = \sup_{\xb, \yb \in \Omega} \| \xb - \yb\|,
\end{equation}
the diameter on this compact space.
Because it is impractical to directly specify the target distribution $\rho_*:\Omega\to \RR$ for a high-dimensional interacting particle system, we instead specify the target value of a given observable $f:\RR^d\to \RR$ and denote this target value by $f_*.$
For technical reasons described below, we assume that $f$ is Lipschitz continuous (essentially meaning that its derivative remains bounded) with Lipschitz constant $K=1$; any value of $K$ could be used and consequently the bound derived below will have a prefactor of $C K.$
We assume that the dynamics of system is governed by an overdamped Langevin equation controlled by an external force $u$; that is, 
\begin{equation}
    d\Xb_t^u = [b(\Xb_t^u) + u(\Xb_t^u)]dt + \sqrt{2D} d\Wb_t,
    \label{eq:appusde}
\end{equation}
as discussed in the main text. 
We also assume that the resulting dynamics is ergodic so that the process \eqref{eq:appusde} relaxes to a unique stationary distribution $\rho_u.$

Naturally, the objective function for the controller seeks to match the time-averaged value of the observable $f$ with the target value $f_*$.
That is, we want to find the optimal controller $u_*$ that solves 
\begin{equation}
    u_* = \argmin_u |\EE_{u} f - f_*|
\end{equation}
over all control functions $u.$
Here the notation $\EE_u$ denotes an expectation 
\begin{equation}
    \EE_u f = \int_{\Omega} f(\xb) \rho^u_{\rm ss}(\xb) d\xb = \lim_{T\to\infty} \frac1T \int_0^T f(\Xb_t^u) dt,
\end{equation}
where $\Xb_t^u$ solves \eqref{eq:appusde}.
Though we do not know functional form of the target steady state density, we assume that it is ``implementable'' in the sense that 
\begin{equation}
    f_* = \EE_* f = \lim_{T\to\infty} \frac1T \int_0^T f(\Xb_t^*) dt,
\end{equation}
where 
\begin{equation}
    d\Xb_t^* = b_*(\Xb_t^*)dt + \sqrt{2D} d\Wb_t,
    \label{eq:targetsde}
\end{equation}
for some unknown $b_*.$

In the approach we take, the optimal controller depends on the choice of observable.
Some choices of $f$ may not be particularly informative about the system---for example, the bulk density could remain fixed for steady state distributions with differing microscopic structures. However, issues with uninformative observables could be overcome by allowing any 1-Lipschitz function $g:\Omega\to \RR$ and carrying out the ``adversarial'' optimization
\begin{equation}
    \max_g \min_u | \EE_u g - \EE_* g |
\end{equation}
so that the optimal controller must minimize the mean discrepancy for \emph{every} observable. 
This stringent requirement actually amounts to minimizing a measure of the distance between the probability distribution with density $\rho_u$ and the unspecified target distribution $\rho_*.$
In fact, the maximization coincides with the Kantorovich-Rubenstein dual formulation of the Wasserstein-1 optimal transport distance~\cite{villani_topics_2003},
\begin{equation}
    \mathcal{W}_1(\rho_u, \rho_*) = \inf_{\pi \in \Pi(\rho_u,\rho_*)} \int_{\Omega\times \Omega} |\xb-\yb| d\pi(\xb,\yb)
    \label{eq:w1}
\end{equation}
where $\Pi(\rho_u, \rho_*)$ denotes all distributions with marginals $\rho_u$ and $\rho_*.$ 
This so-called integral probability metric is related to other measures of ``distance'' between probability distribution~\cite{gibbs_choosing_2002} and is upper bounded by the total variation distance
\begin{equation}
     \mathcal{W}_1(\rho_u, \rho_*) \leq C\| \rho_u - \rho_* \|_{\rm TV} 
\end{equation}
which is proved by using the coupling definition of the total variation distance and noting that the expected distance on the coupling that realizes the infimum in~\eqref{eq:w1} must be less than the maximum distance between points in $\Omega$ multiplied by the expected fraction of unequal points.
Via Pinsker's inequality, this bound can be extended to the KL divergence between the two distributions:
\begin{equation}
     \mathcal{W}_1(\rho_u, \rho_*) \leq C\| \rho_u - \rho_* \|_{\rm TV} ,
                                   \leq C \sqrt{D_{\rm KL}(\rho_u\| \rho_*)}. 
\end{equation}

The KL divergence, or relative entropy, is a well-studied object in nonequilibrium statistical mechanics because it provides a physical measure of entropic distance between distributions~\cite{sivak_nearequilibrium_2012}. 
The steady state distributions appearing in \eqref{eq:klineq} differ with an entropic cost that can be bounded by considering the time average of the relative entropy of the path measures associated with the corresponding nonequilibrium trajectories of duration $T$~\cite{pantazis_relative_2013},
\begin{equation}
  \begin{aligned}
    D_{\rm KL}(\rho_u \| \rho_*) &\leq \frac{1}{T} D_{\rm KL}(\mathbb{P}_u \| \mathbb{P}_*)\\
    &= \frac1T \EE_{\mathbb{P}_u} \log \frac{d\mathbb{P}_u}{d\mathbb{P}_*}, \\
    &= \EE_{\mathbb{P}_u} \frac{1}{\sqrt{2D}T}\int_0^T \delta u(\Xb_t) d\Wb_t + \frac1{2DT} \int_0^T \| \delta u(\Xb_t) \|^2 dt.
    \end{aligned}
\end{equation}
In the expressions above $\delta u = b_* - (b+u)$.
The stochastic integral vanishes in the limit $T\to\infty$ and we see that the obvious minimizer can be deduced:
\begin{equation}
     \EE_{\mathbb{P}_u} \frac1{2DT} \int_0^T \| b_*(X_t) - b(X_t) - u(X_t) \|^2 dt \implies u_*(X_t) = b_*(X_t) - b(X_t)
\end{equation}
where $u_* = \argmin_u D_{\rm KL}(\mathbb{P}_u \| \mathbb{P}_*),$ which matches the target drift function exactly.
However, there is an entropic cost associated with this choice compared to the path measure $\mathbb{P}$ of the unperturbed dynamics;
\begin{equation}
    D_{\rm KL}(\mathbb{P}_u \| \mathbb{P}) =  \frac1{2DT} \EE_{\mathbb{P}_u} \int_0^T \| u(\Xb_t^u) \|^2 dt \equiv \sigma_{\rm ex} \geq 0
\end{equation}
is non-negative excess entropy associated with the control.
This quantity vanishes when $u\equiv 0$ meaning that no control is required to produce the steady state. 
In the long time limit, it could also vanish if $u = \nabla V$, for some potential energy $V$, is a conservative force, the ``interaction design'' paradigm discussed in the introduction---in this case any excess dissipation would only be transient. 
We conclude that for the protocol $u$ that results from optimization (via reinforcement learning or some other technique) will dissipate at least $\sigma_{\rm ex}$ in order match the true target distribution. 

\section{Computational details of for Deep Q-Learning}\label{app:dql}

\paragraph*{Code Availability:}
All simulations were conducted using OpenMM 7.5.1 and PyTorch 1.9.0.
Our code is available at \url{https://github.com/rotskoff-group/dissipative-design}. 

$Q$-learning is a model-free, off-policy Reinforcement Learning algorithm used to estimate the optimal state-action value function $Q^{\star}$. 
Because $Q$-learning is model-free we can avoid incorporating the dynamics of the system into the optimization framework, which may be unknown in experimental systems.

In many RL algorithms, we assume that the environment is a Markov decision process; in our case, this means we consider a dynamics specified by states $\xb\in \RR^d$, actions of the external protocol $\ab \in [0, a_{\rm max}]^{n_{\rm regions}}$, and a cost function $\mathcal{C}.$
In deep $Q$-learning\cite{mnih-atari-2013}, we represent the state-action value function, $Q$, using a deep neural network, and optimize this network using the Bellman dynamic programming principle. 
For a deterministic policy $u$, the Bellman equation reads
\begin{equation}
\label{eq:appQ}
    Q^u(\xb_t, \ab_t) = \EE^{\xb_t}_u \left[ \mathcal{C}(\xb_{t+\tau}; \ab_t) + \sum_{k=1}^\infty \gamma^k \mathcal{C}(\xb_{t+\tau(k+1)}) \right].
\end{equation}
In this expression, the cost is evaluated for a state $\xb_{t+\tau}$ after taking action $\ab_t$ and subsequently following the protocol $u.$
In practice, we use a second target neural $Q^{\prime}$ network to estimate the predicted value of $Q^u(\xb_{t + \tau}, u(\xb_{t + \tau}))$ to ensure more stable updates.
We update the weights of the target network to match the weights of the $Q$ network at a rate $\tau$. 
Finally, we use experience replay, a technique in which we store past experiences in a replay buffer $\mathcal{R}$ and sample from the buffer during update steps. Each experience here is a tuple of ${(\xb_t, \ab_t, c_t, \xb_{t+1})}$ that was observed within each ``grid'' (i.e. a spatial region) of control.

Our state space $\mathcal{S}$ represents the normalized empirical cluster-size distribution over some sample space $\Omega$, which was specified differently for each of our target distributions, as described in the main text.
We found that this representation worked the best compared to other possible state representations, such as images of the system configurations. Each state $\xb^{i}_t$ was the normalized cluster-size distribution within a grid of the system $\Xb_t$. This allowed us to utilize local information about the system to spatially control the system.

$Q$-Learning methods are known to suffer from underestimation bias~\cite{fujimoto_addressing_2018}. 
This bias arises during training because we are using $\min_{\ab}{Q^u(\xb, \ab)}$ as an estimate for $\min_{\ab}{Q^{\star}(\xb, \ab)}$. By Jensen's Inequality, we have
\begin{equation}
{\EE{\min_{\ab}{Q(\xb, \ab)}} \le {\min_{\ab}\EE{Q(\xb, \ab)}} =  \min_{\ab}{Q^{\star}(\xb, \ab)}}   
\end{equation}
 Underestimation bias results in suboptimal actions, which will have an artificially lower state-action value, being selected as the optimal action. 
 One approach to mitigating this bias is to use clipped double $Q$-learning, where we maintain two estimates of $Q^{\star}$: $Q_{1}$ and $Q_{2}$~\cite{fujimoto_addressing_2018}. We then use the maximum of the two estimates provided by these networks as an upper bound of the estimate of the state-action value. With this estimate, we update both $Q_{1}$ and $Q_{2}$. 
 Finally, we update the weights of the target networks $Q_{1}^{\prime}$ and $Q_{2}^{\prime}$ towards the weights of $Q_{1}$ and $Q_{2}$ at a rate $\tau$.
 
Reinforcement Learning algorithms are constrained by a trade-off between exploration and exploitation~\cite{sutton_reinforcement_2018}. In order to determine an optimal policy $\pi,$ it is necessary to \emph{exploit} our current estimates of the state-action value. 
However, it is necessary to sufficiently \emph{explore} the state-action space in order to improve our current estimates of the state-action value. 
One common approach to this dilemma is to use an $\epsilon$-greedy search, where a random action is selected with probability $\epsilon$ during training to promote exploration. In our approach, we instead always randomly select an action during the first $e_{\rm explore}$ episodes of training. After the first $e_{\rm explore}$ episodes, we decided to train our state-action value function using a greedy approach (i.e. selecting the optimal action that minimizes the cost). We used this approach because of the small size of our action space $\mathcal{A}$ and because the decisions made in each grid were considered to be independent experiences.
This allowed us to more thoroughly explore the action-space during the first $e_{\rm explore}$ episodes, especially when using higher-resolutions of control. 

When using higher-resolutions of control, it is possible for a grid to not contain any particles. During training, if a grid did not contain any particles at time $t$ or $t + 1$, it was not included in the replay buffer $\mathcal{R}$. While training with the greedy approach (i.e. if the episode $e \ge e_{\rm explore}$), we set the action to be $a_{\rm max}$ if the grid did not contain any particles at some time $t$. For our Lennard-Jones system, $a_{\rm max}$ represented the highest possible temperature, in order to ensure that any free particle that entered the region would continue to diffuse until it reached a region with more particles. For our active matter system $a_{\rm max}$ represented the highest possible activity, to similarly ensure that any free particle that entered the region would continue self-propulsion until it was able to form clusters. 
Finally, when a cluster was located in multiple grids, each of these grids was considered to contain the entire cluster. This effectively allows neighboring grids that shared a cluster to ``communicate,'' which was especially important during the high-resolution control cases.

\begin{algorithm}[H]
\caption{Clipped Double Q Learning Training\label{alg:cdql}}   
\begin{algorithmic}[1]
    \State {\bfseries Initialize Replay Buffer $\mathcal{R}$ to capacity $\mathcal{N}$} 
    \State {\bfseries Randomly initialize $Q_1$ and $Q_2$ networks with weights ${\theta^{Q_1}}$ and ${\theta^{Q_2}}$}
    \State {\bfseries Initialize target networks $Q_1^{\prime}$ and $Q_2^{\prime}$ networks with weights ${\theta^{Q_1^{\prime}} \leftarrow \theta^{Q_1}}$ and ${\theta^{Q_2^{\prime}} \leftarrow \theta^{Q_2}}$}
    \For {e = $0$ ${\dots}$ $\Mb$} 
    \State Initialize state ${\Xb_0}$ for episode e
        \For {t = $0$ ${\dots}$ $\Tb$}
        \ForEach{$\xb_t$ in ${\Xb_t}$}
            \If{$\xb_t$  is empty and e ${\ge e_{\rm explore}}$}
                \State{Select  ${a_t = a_{\rm max}}$ and execute action ${a_t}$}
                \State{\textbf{continue}}
            \EndIf
            \If {e ${< e_{\rm explore}}$}
            \State{Select a random action ${a_t}$ from $\mathcal{A}$}
            \Else
            \State{Select ${a_t = \argmin_{\ab} Q_1(\xb_t, \ab)}$}
            \EndIf
            \State{Execute action ${a_t}$ and observe next state ${x_{t+1}}$ and cost $c_t \equiv \mathcal{C}$}
             \If{$\xb_{t}$ or $\xb_{t+1}$ is empty}
                \State{\textbf{continue}}
            \EndIf
            \State{Store transition ${(\xb_t, \ab_t, c_t, \xb
            _{t+1})}$ in $\mathcal{R}$}
            \EndFor
        \State{Sample random batch of transitions ${(\xb_j, \ab_j, c_j, \xb_{j+1})}$ from $\mathcal{R}$}
        \State{Set ${y_j = c_j + \gamma \max_{1, 2}\min_{\ab^{\prime}} Q_{1, 2}^{\prime}(\xb_{t + 1}, \ab)}$}
    \State{Update ${Q_{1, 2}}$ by minimizing ${L_{1, 2} = (y_j - {Q_{1, 2}(\xb_j, \ab_j)})^{2}}$}
        \State{Update the target networks ${\theta^{Q_{1, 2}^{\prime}} \leftarrow (1 - \tau){\theta^{Q_{1, 2}^{\prime}}} + \tau{\theta^{Q_{1, 2}}}}$}
            
        \EndFor
    \EndFor
\end{algorithmic}
\end{algorithm}

\begin{table}
\centering
\begin{tabular}{|p{4cm}||p{5cm}|p{3cm}|}
 \hline
 \multicolumn{3}{|c|}{Hyper-Parameters} \\
 \hline
 Hyperparameter &Active Colloids&LJ System\\
 \hline
 $Q$ learning rate  & $3 \times 10^{-4}$    &$3 \times 10^{-4}$\\
 Optimizer&   Adam  & Adam\\
 Target Update Rate ($\tau$) & $5 \times 10^{-3}$ & $5 \times 10^{-3}$\\
 Batch Size    &32 & 32\\
 Discount Factor ($\gamma$)&   0.9  & 0.95\\
 Number of Hidden Layers & 2  & 1\\
$\eb_{\rm explore}$ & 5  & 25\\
$\Tb$ & 350  & 150\\
$a_{high}$ & 1.5 & 1.0\\
$\mathcal{A}$ & $[0, 0.25, 0.5, 0.75, 1.0, 1.25, 1.5]$ & $[0.01, 0.25, 1.0]$\\ 
 \hline
\end{tabular}
\caption{Relevant parameters for $Q$-learning training}
\end{table}

\section{Active Colloids}\label{app:ac}

We modeled our active colloids based on the experimental system in ~\cite{palacci_living_2013}. In this system, colloidal activity can be modulated by blue light. In addition to self-propulsion, these particles exhibit an attractive interaction due to phoretic and osmotic effects when activated by blue light.

\begin{figure}
    \centering
    \includegraphics[width=0.45\linewidth]{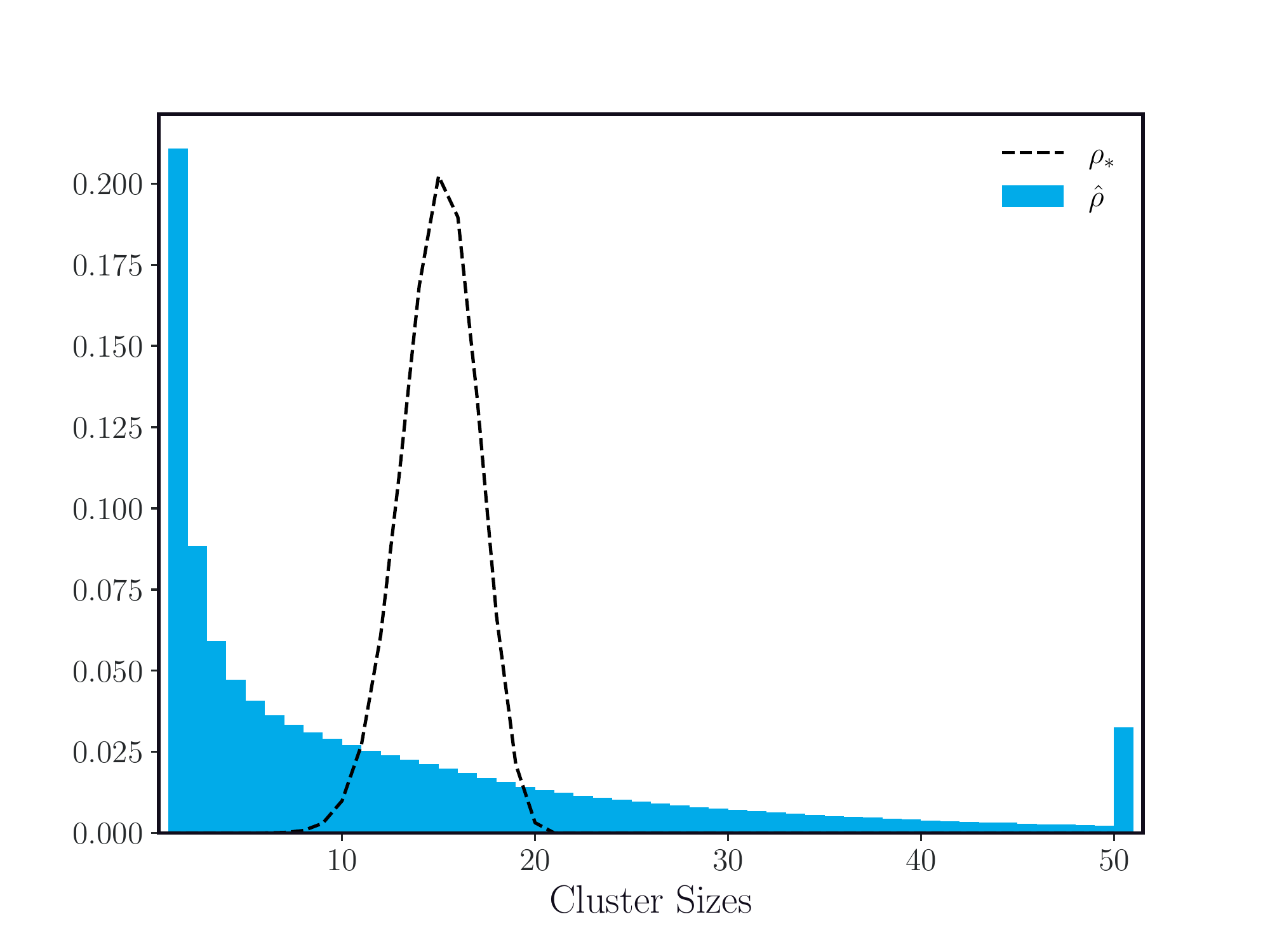}
    \includegraphics[width=0.45\linewidth]{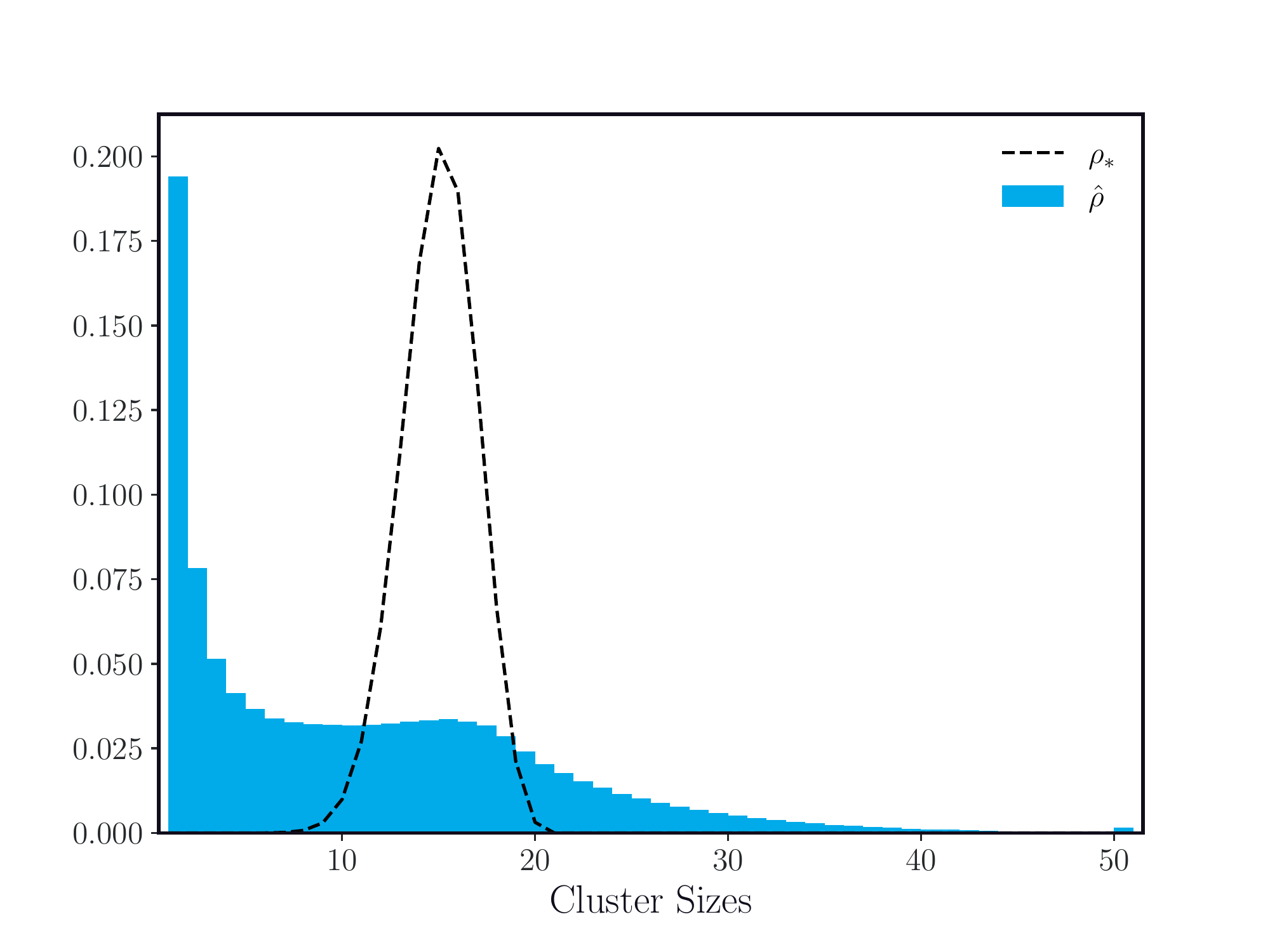}
    \includegraphics[width=0.45\linewidth]{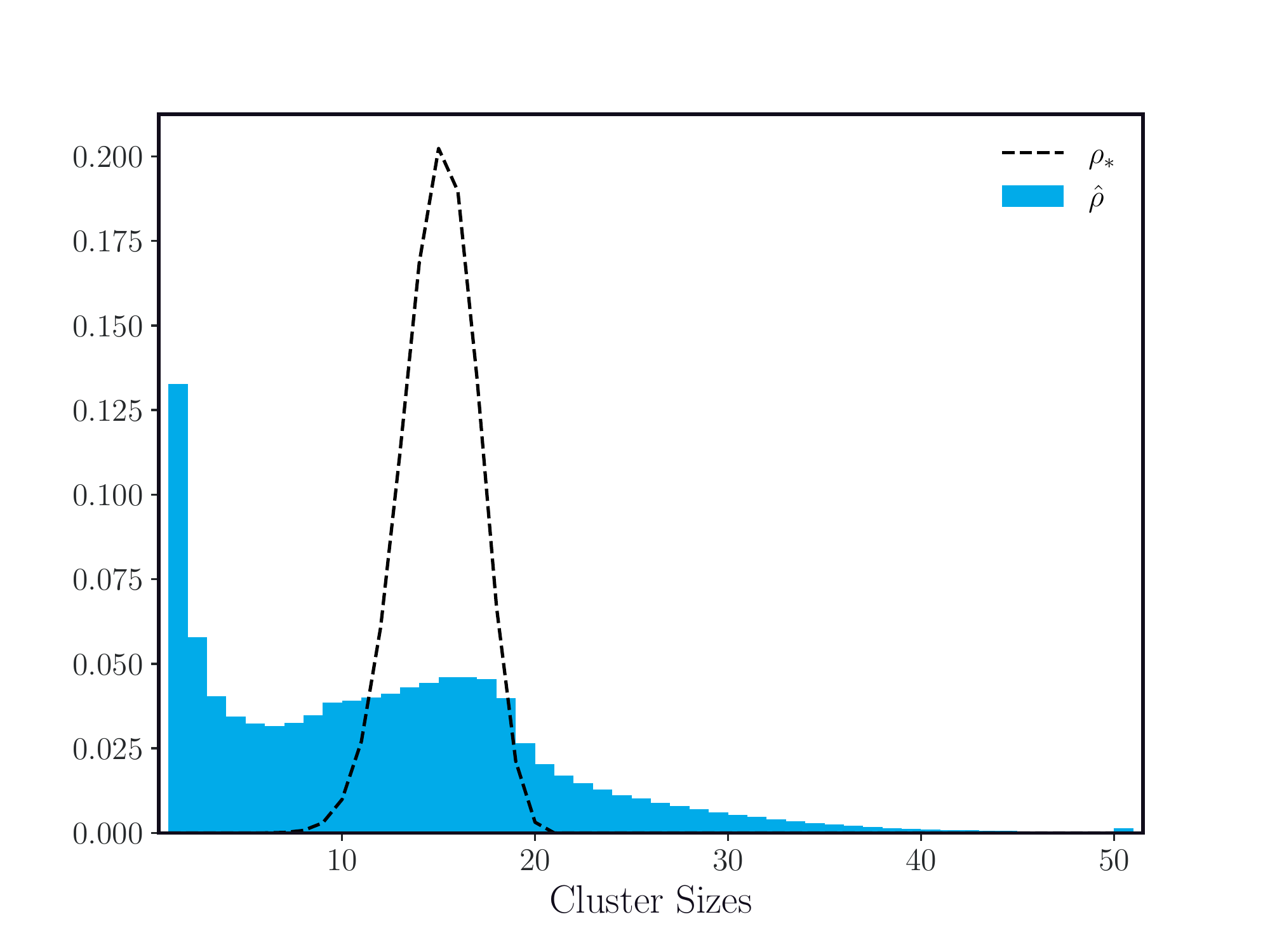}
    \includegraphics[width=0.45\linewidth]{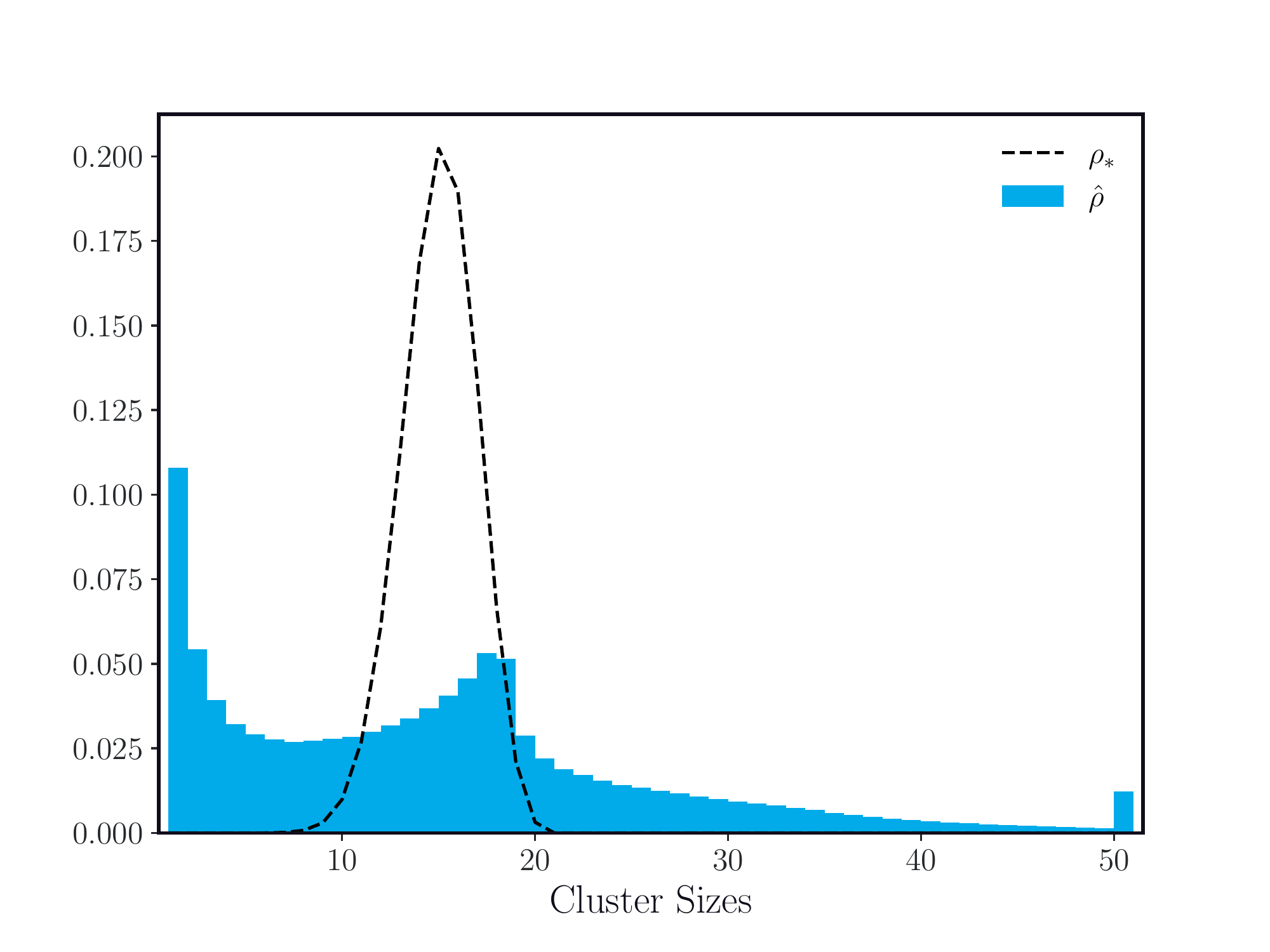}
    \includegraphics[width=0.32\linewidth]{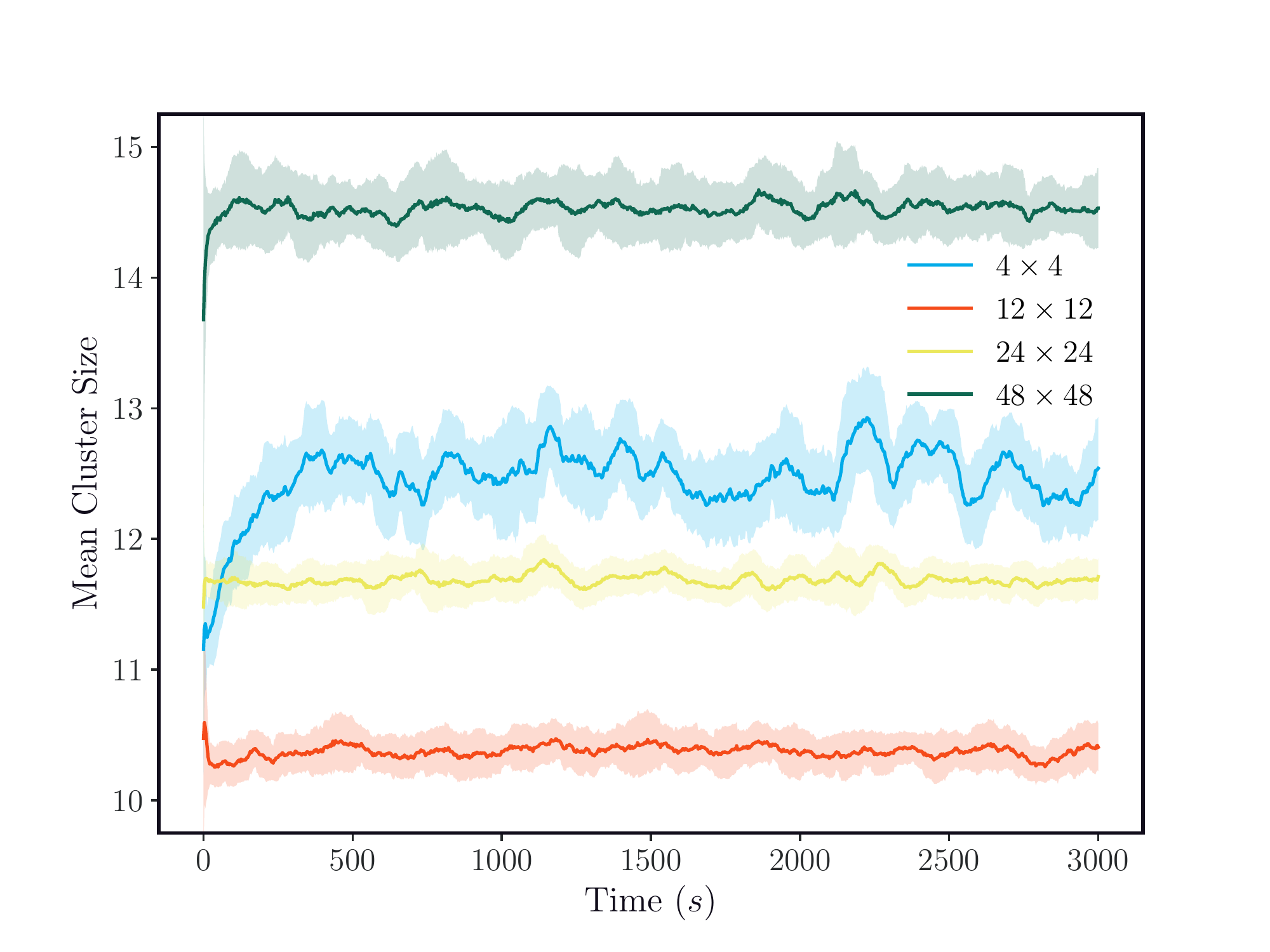}
    \includegraphics[width=0.32\linewidth]{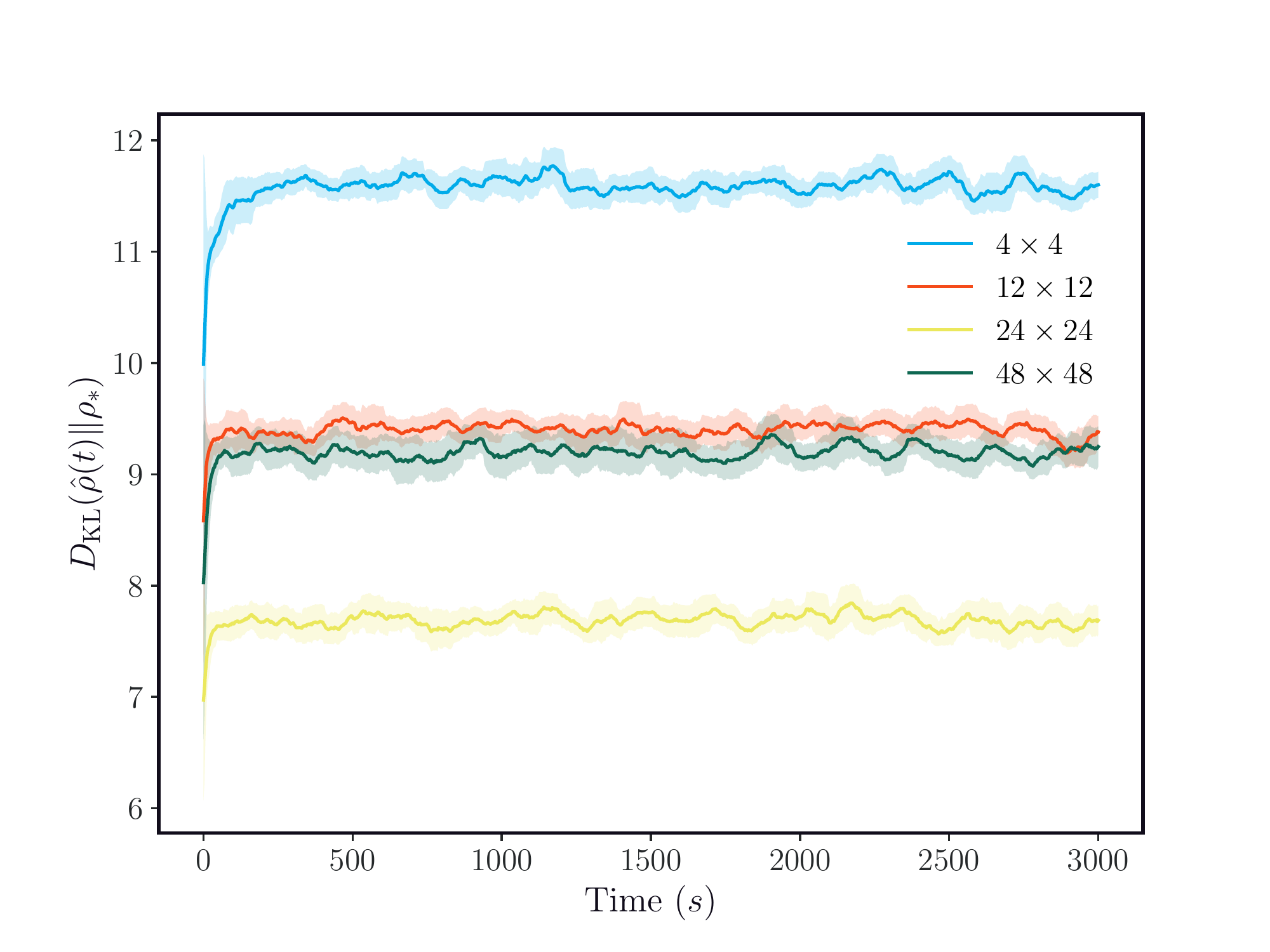}
    \includegraphics[width=0.32\linewidth]{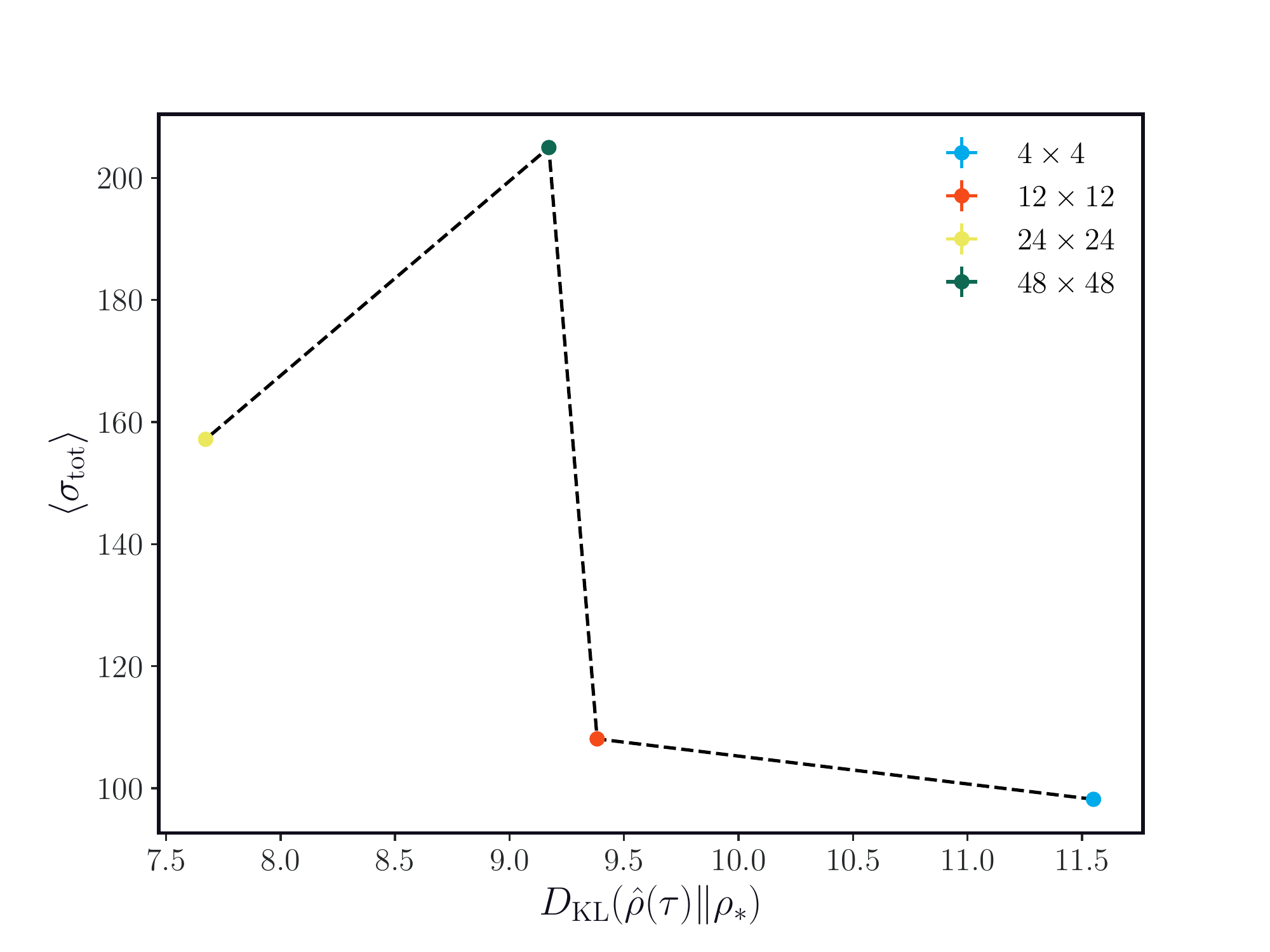}
    \caption{Summary of results for Binomial target distribution. Histograms of cluster sizes for $4\times 4$ (top left) $12\times 12$ (top right) $24\times 24$ (center left) and $48 \times 48$ (center right). Mean cluster sizes (bottom left), KL cost function (bottom center), and total entropy production as a function $D_{\rm KL}$ (bottom right).}
    \label{fig:my_label}
\end{figure}

\begin{figure}
    \centering
    \includegraphics[width=0.45\linewidth]{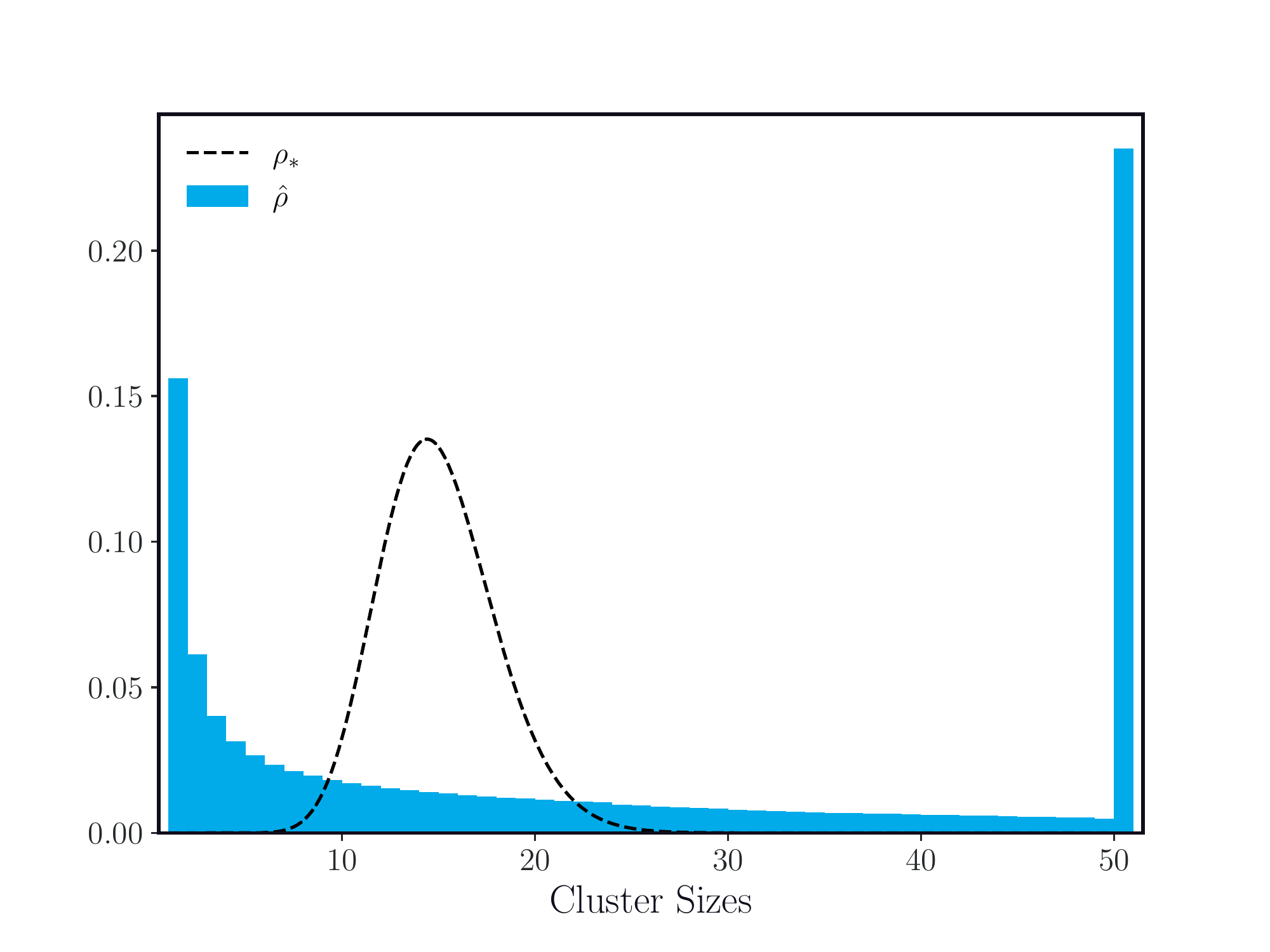}
    \includegraphics[width=0.45\linewidth]{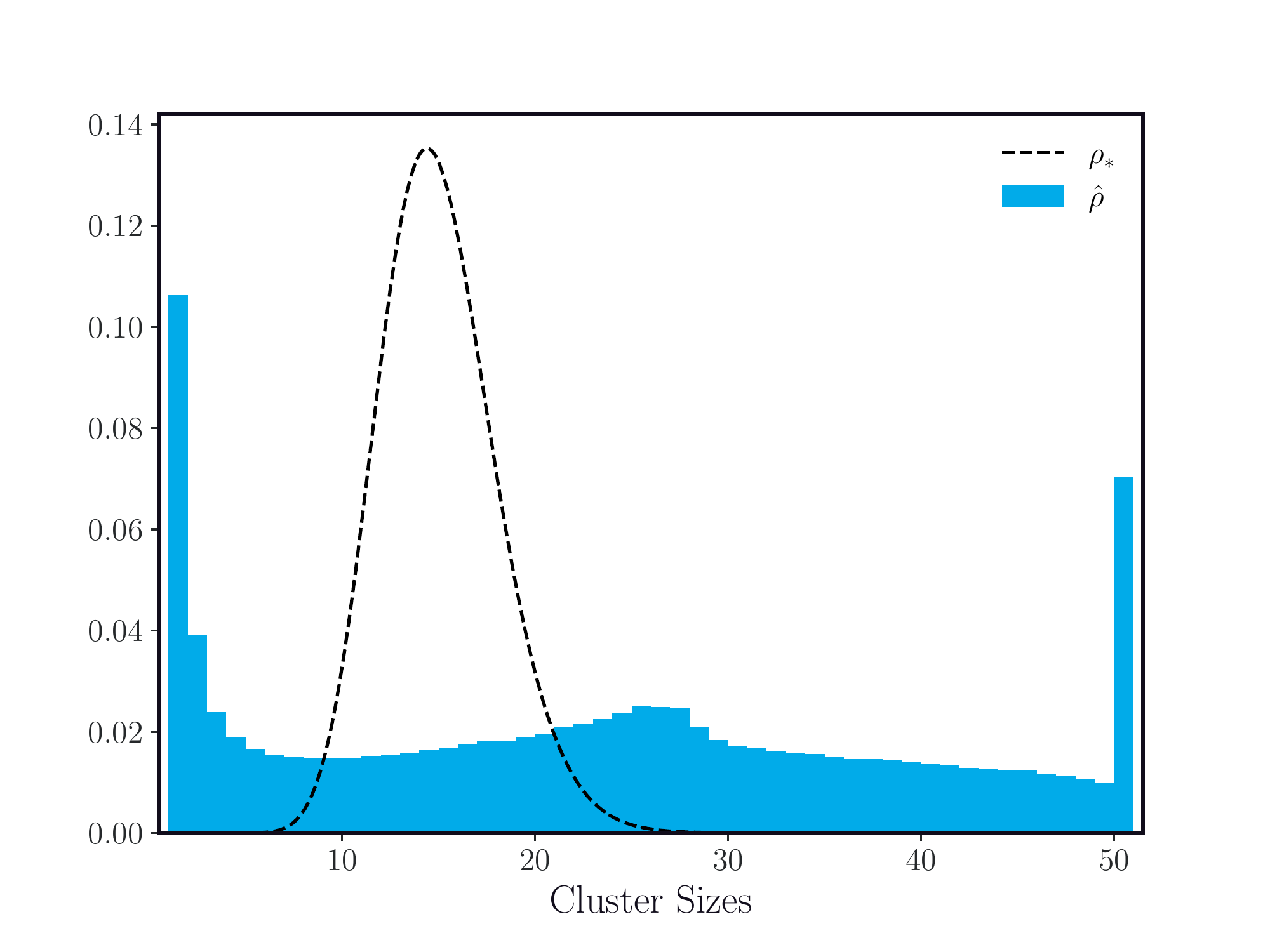}
    \includegraphics[width=0.45\linewidth]{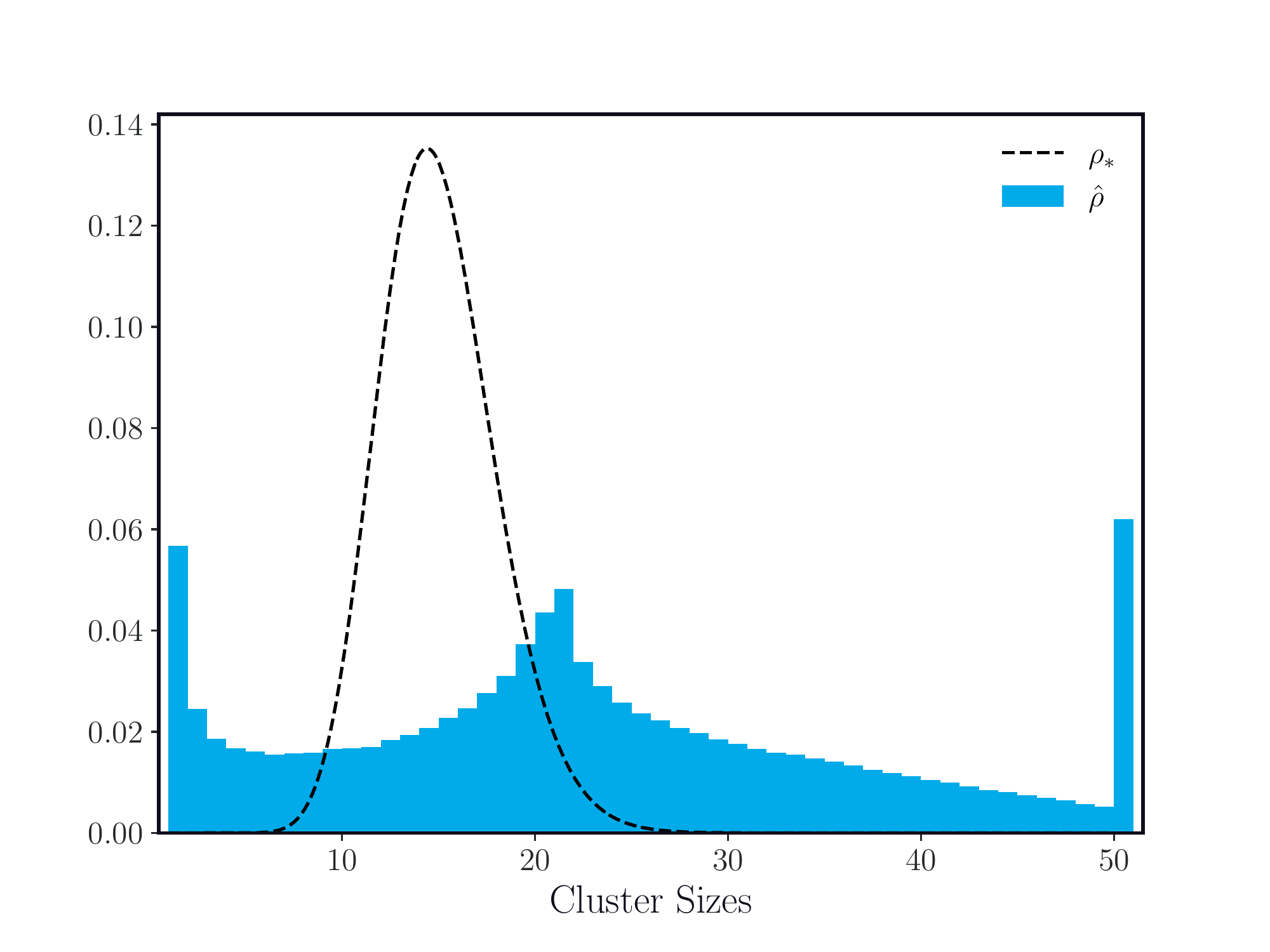}
    \includegraphics[width=0.45\linewidth]{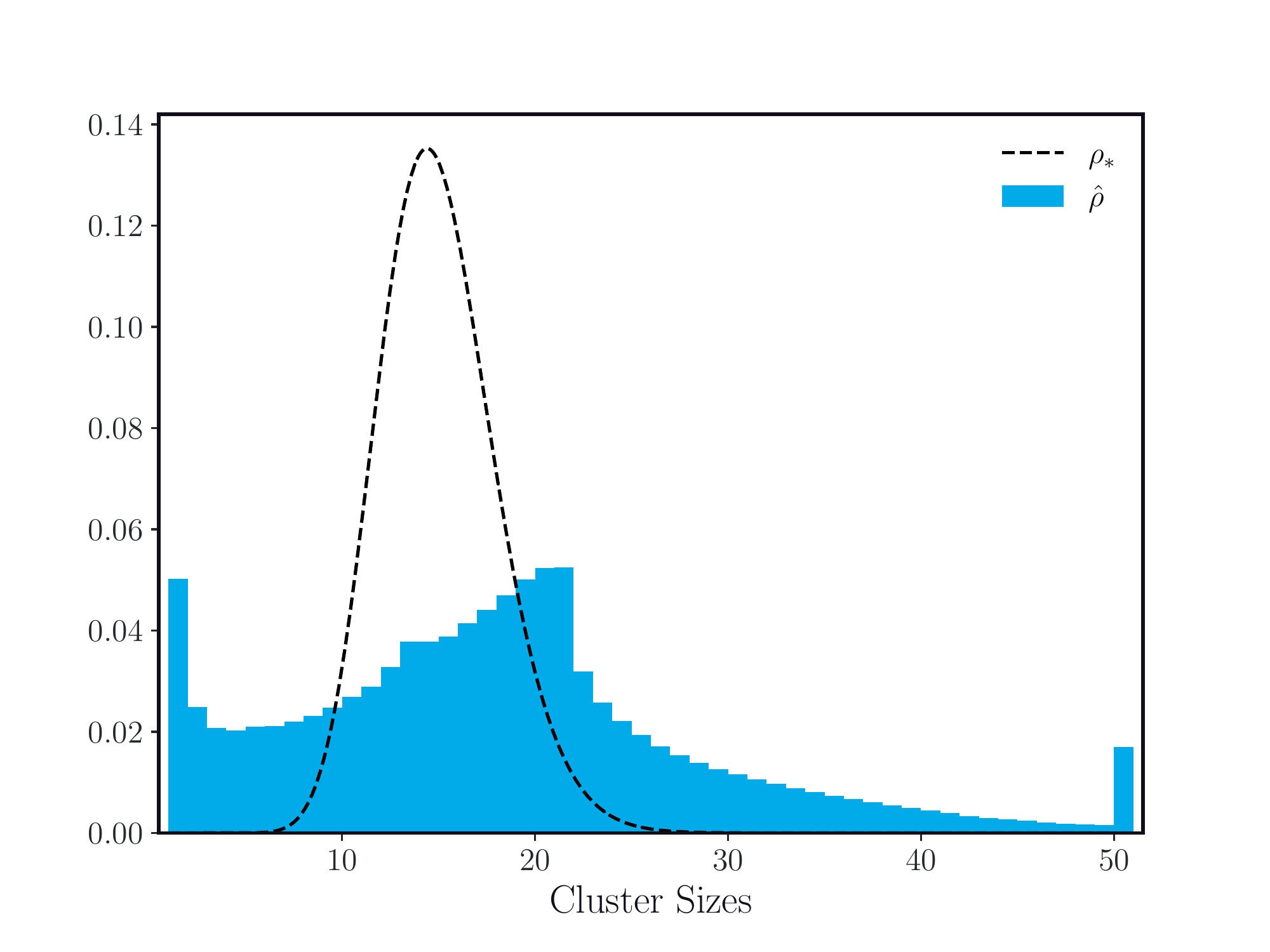}
    \includegraphics[width=0.32\linewidth]{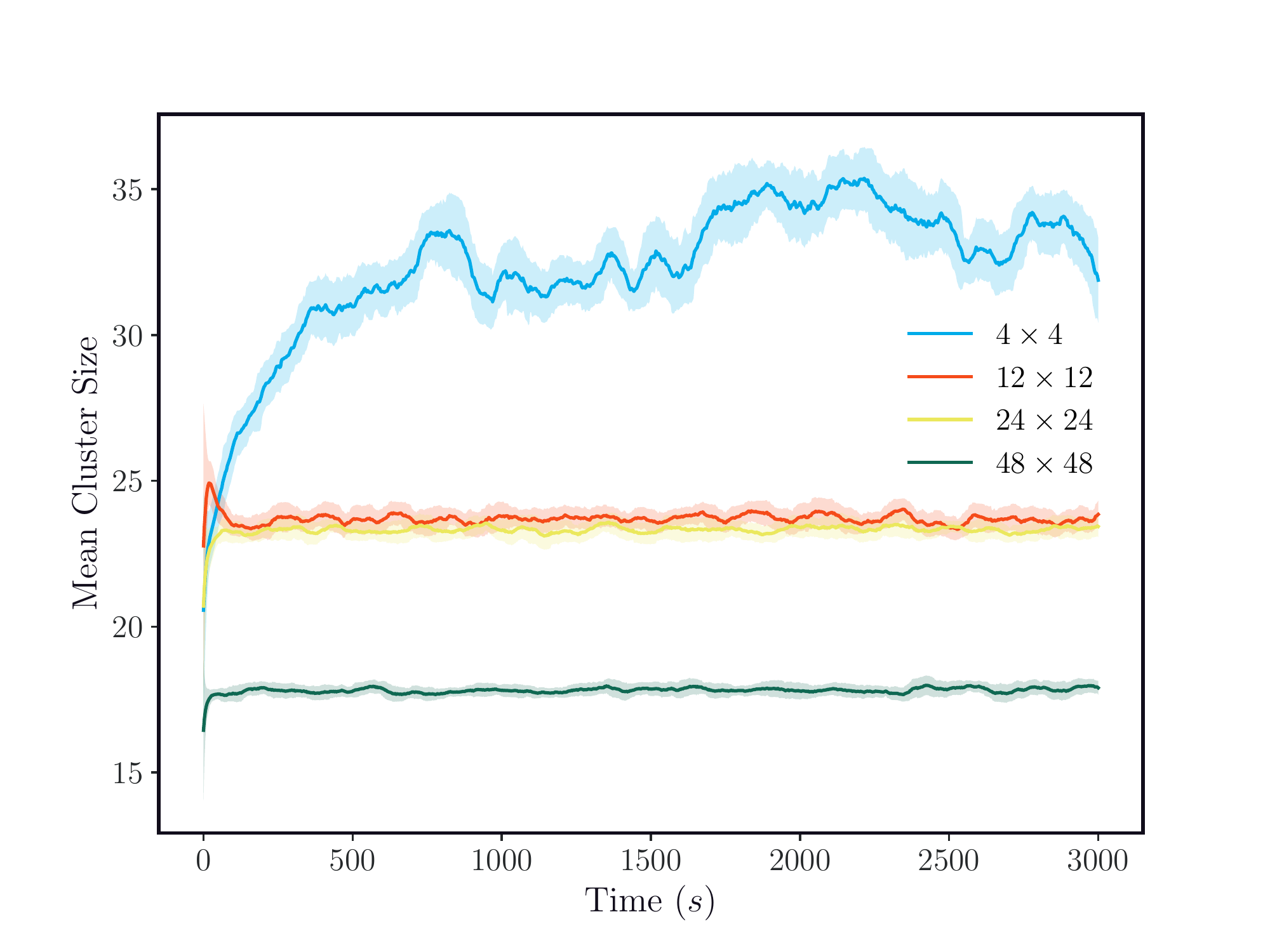}
    \includegraphics[width=0.32\linewidth]{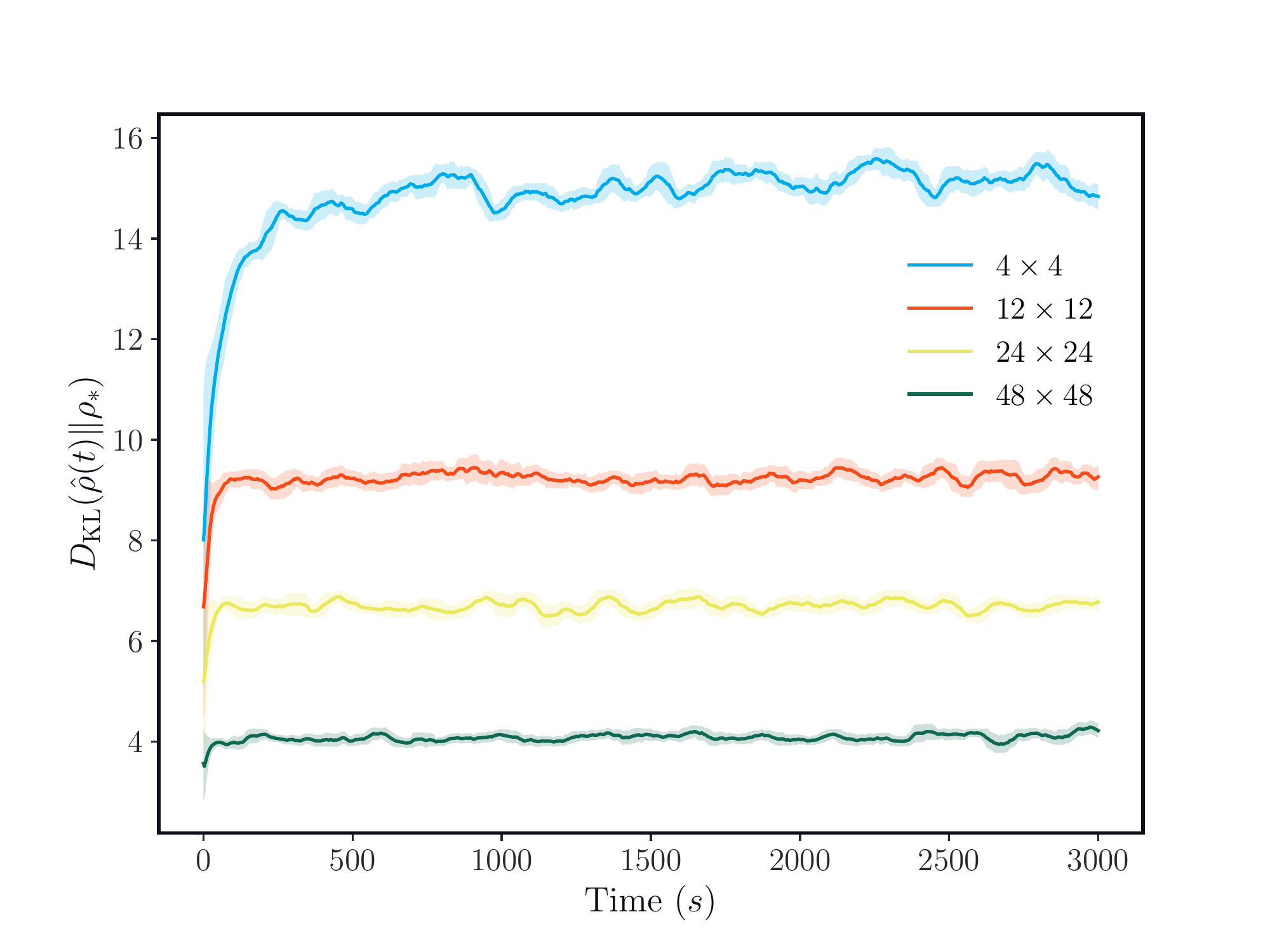}
    \includegraphics[width=0.32\linewidth]{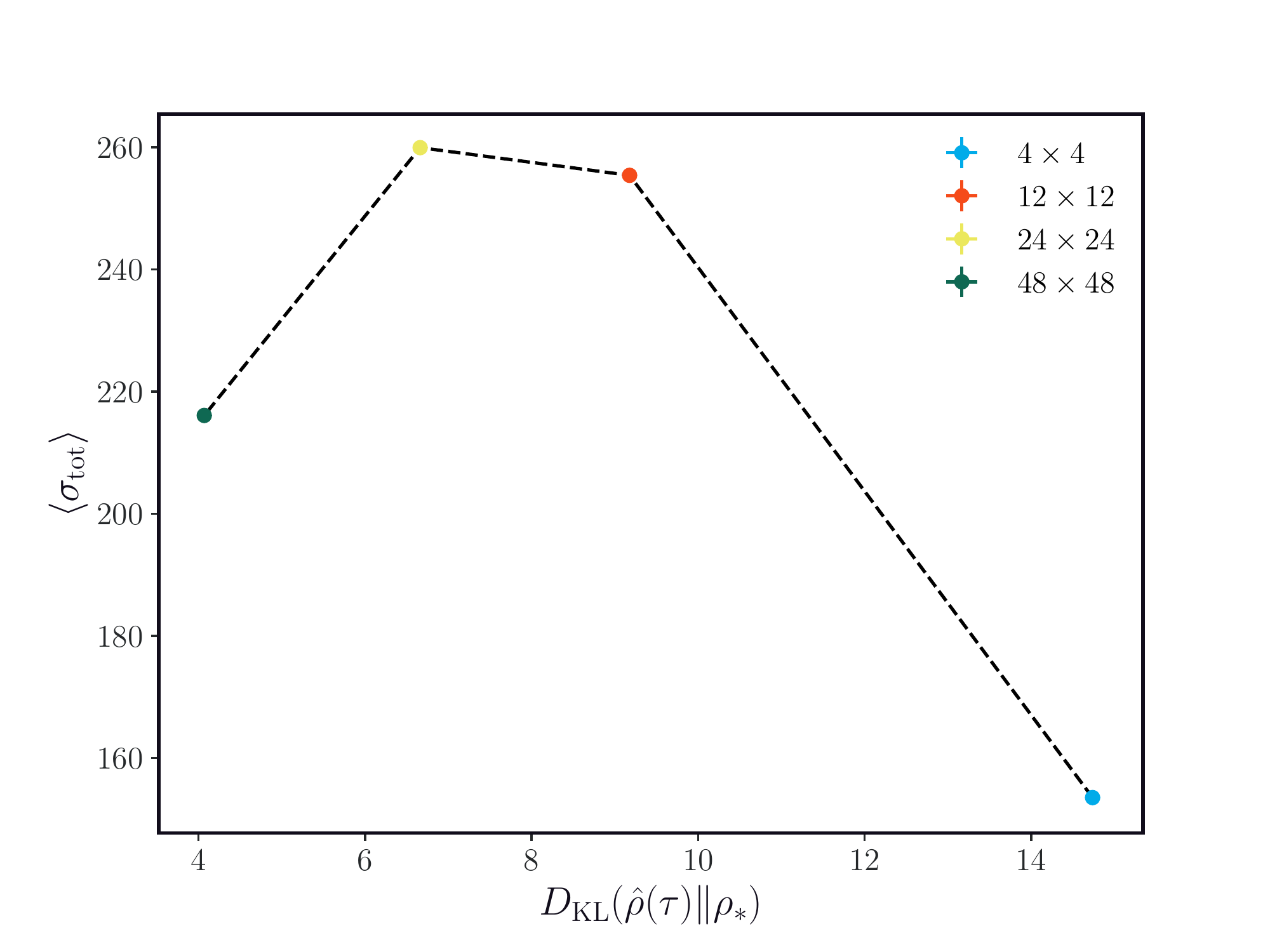}
    \caption{Summary of results for Gamma target distribution. Histograms of cluster sizes for $4\times 4$ (top left) $12\times 12$ (top right) $24\times 24$ (center left) and $48 \times 48$ (center right). Mean cluster sizes (bottom left), KL cost function (bottom center), and total entropy production as a function $D_{\rm KL}$ (bottom right).}
    \label{fig:gammasummary}
\end{figure}

\begin{figure}
    \centering
    \includegraphics[width=0.45\linewidth]{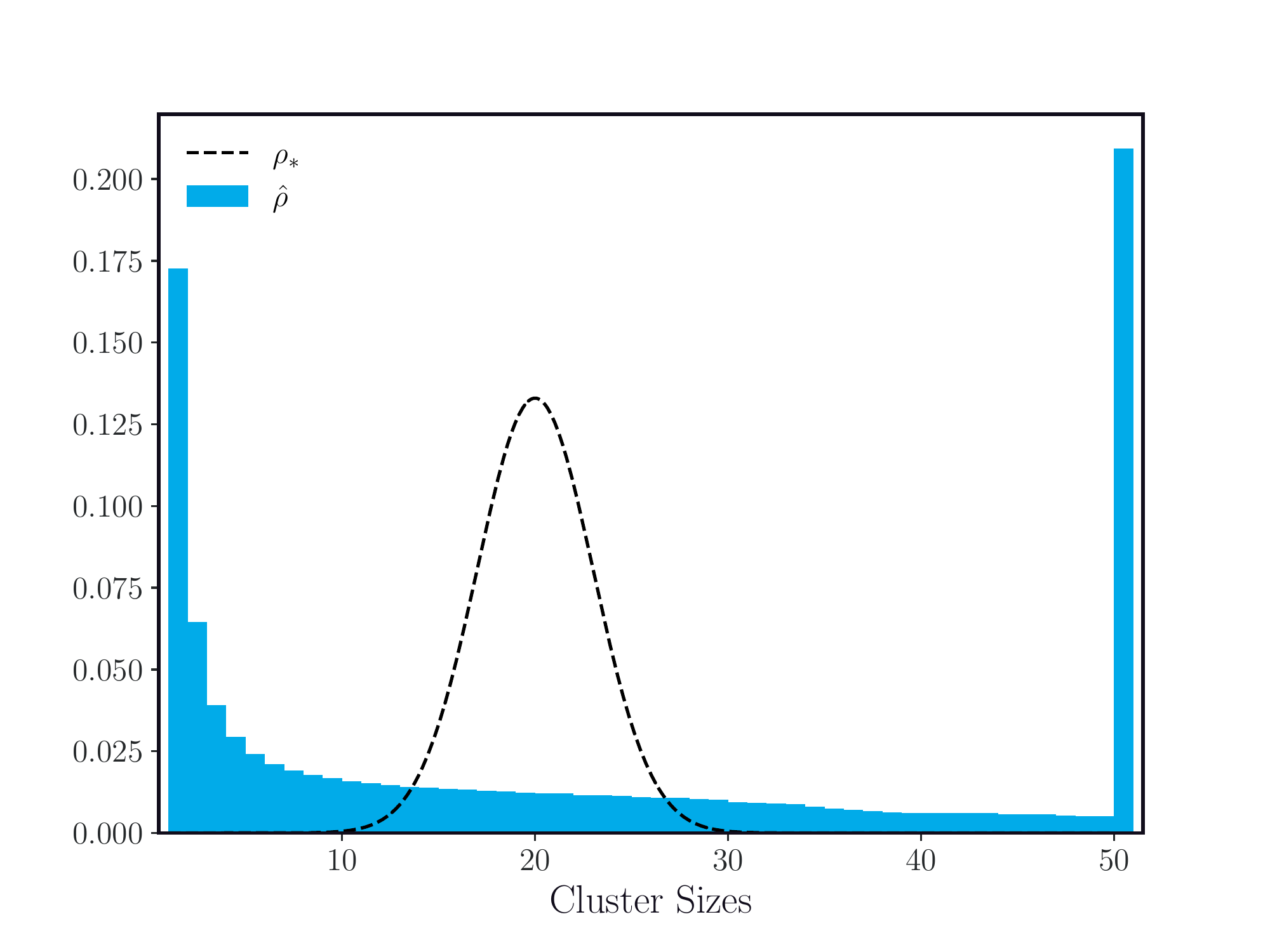}
    \includegraphics[width=0.45\linewidth]{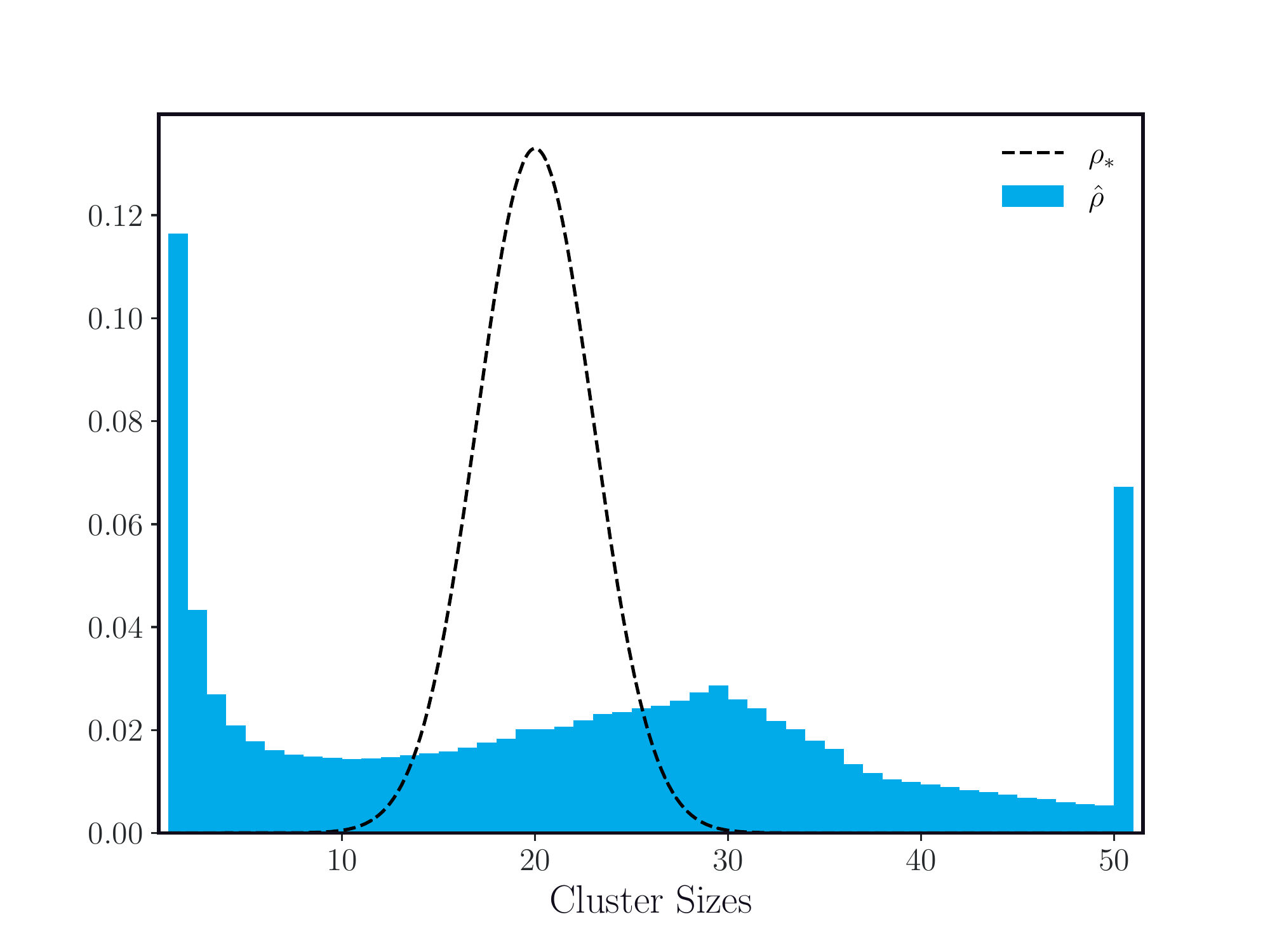}
    \includegraphics[width=0.45\linewidth]{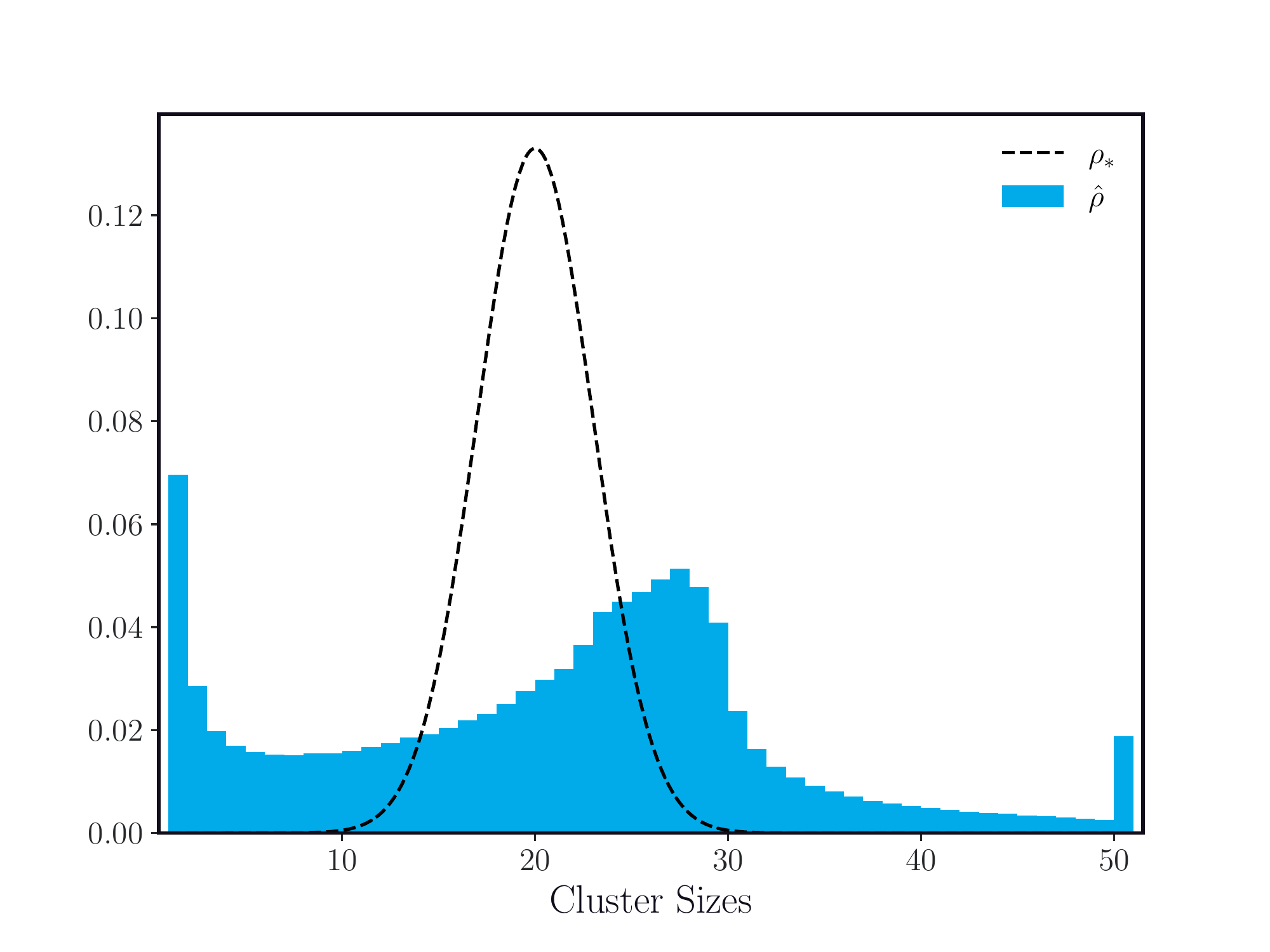}
    \includegraphics[width=0.45\linewidth]{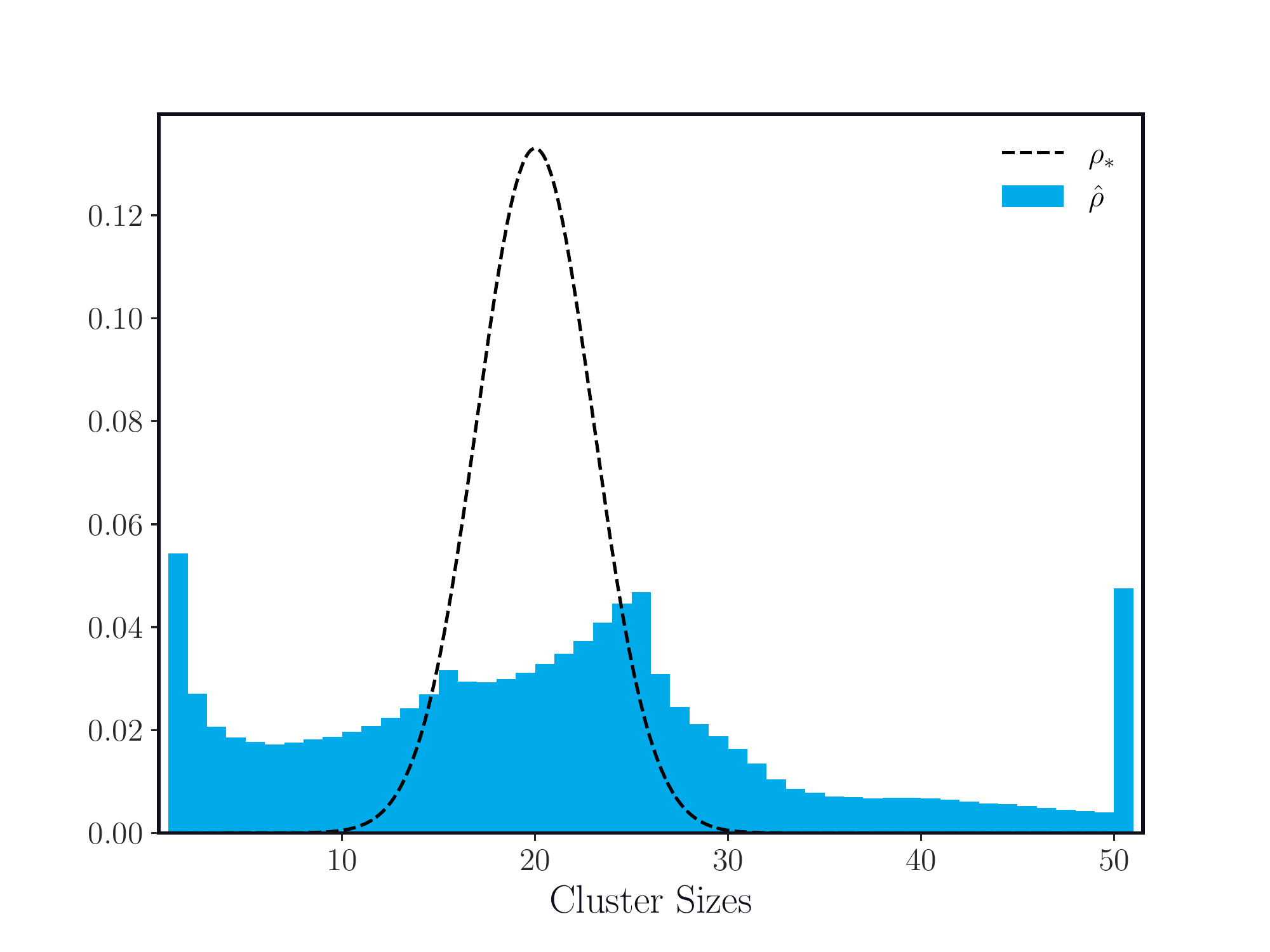}
    \includegraphics[width=0.32\linewidth]{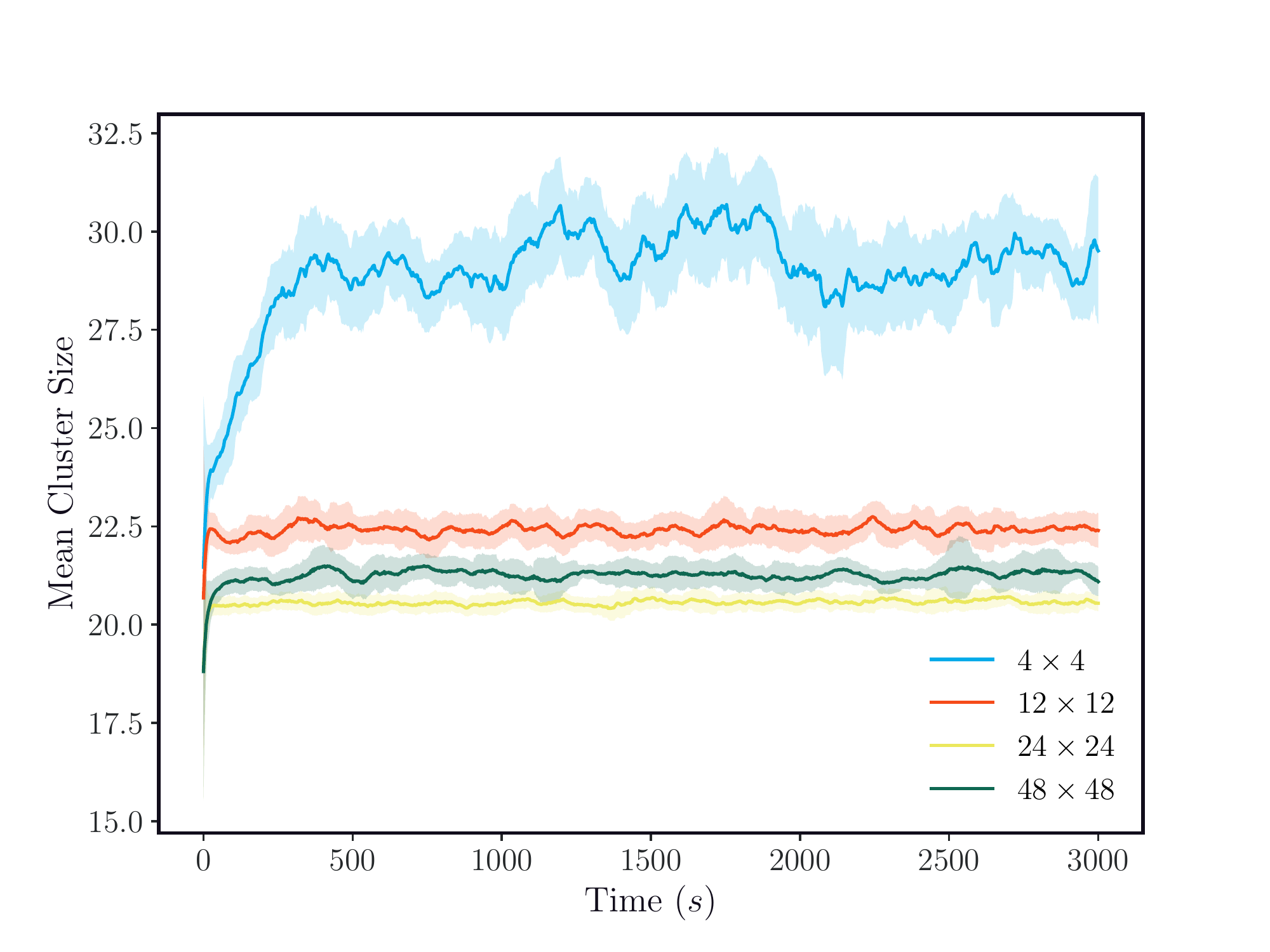}
    \includegraphics[width=0.32\linewidth]{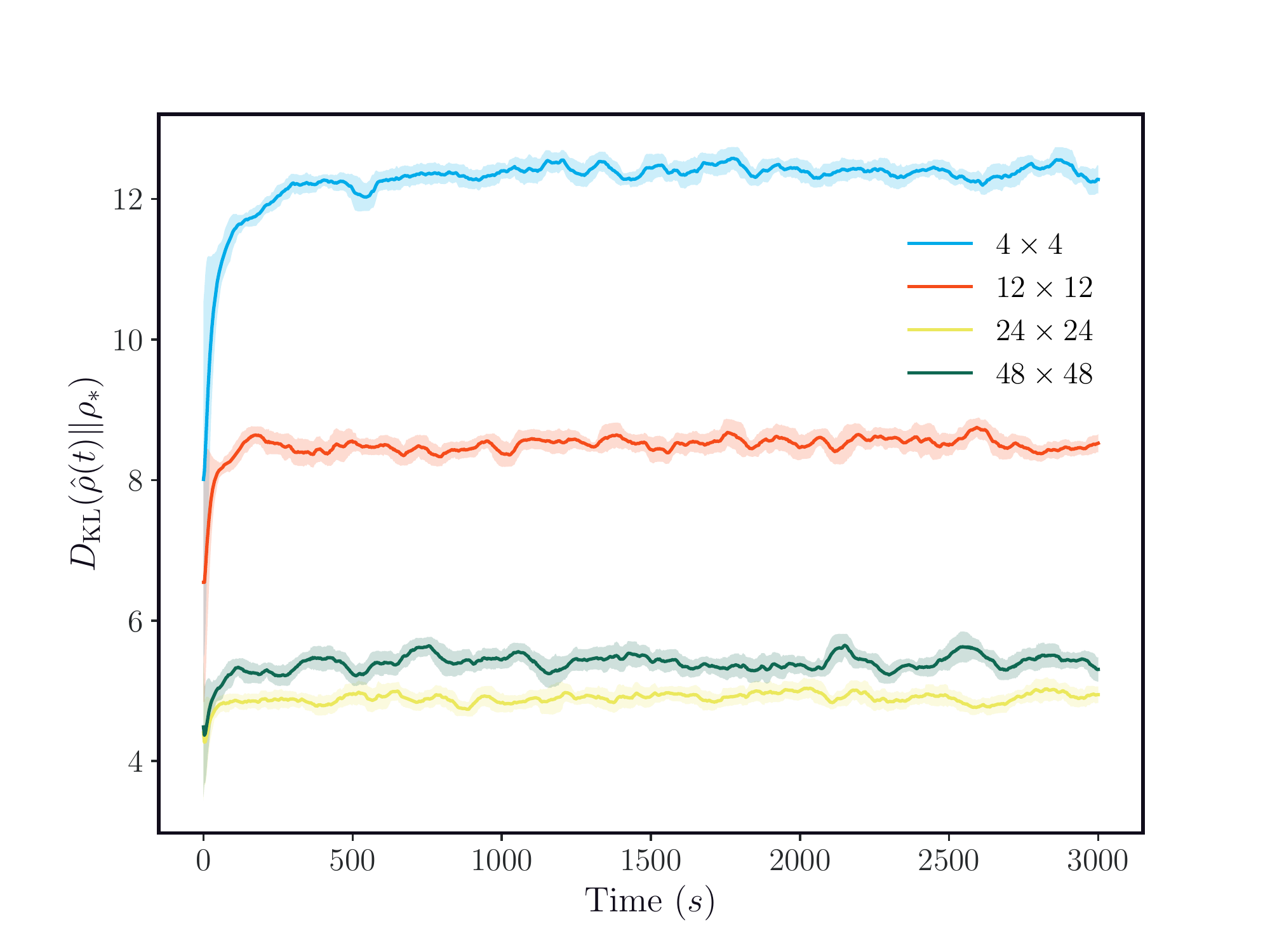}
    \includegraphics[width=0.32\linewidth]{gaussian_sigma_kl.pdf}
    \caption{Summary of results for Gaussian target distribution. Histograms of cluster sizes for $4\times 4$ (top left) $12\times 12$ (top right) $24\times 24$ (center left) and $48 \times 48$ (center right). Mean cluster sizes (bottom left), KL cost function (bottom center), and total entropy production as a function $D_{\rm KL}$ (bottom right).}
    \label{fig:gaussiansummary}
\end{figure}

We work in normalized units of the particle diameter and the self-propulsion velocity.
The model consists of a purely repulsive interaction
\begin{equation}
U_{\rm WCA}(r) = 
\begin{cases}
      0 & r > r_{\rm cut}  \\
  4 \epsilon \left[ \left(\frac{\sigma}{r}\right)^{12} - \left(\frac{\sigma}{r} \right)^{6} \right] + \epsilon   & r \leq r_{\rm cut} 
\end{cases} 
\end{equation}
with $r_{\rm cut} = 2^{1/6}\sigma$.
In addition to the hard sphere repulsion, there is an attractive force induced by the activity which models hydrodynamic effects,
\begin{equation}
U_{\rm attract}(\xb_i, \xb_j) = \frac{\sqrt{A_i A_j}}{\| \xb_i - \xb_j \|^2} 
\end{equation}
where the coefficient $A_i$ is determined by the instantaneous value of the activity
\begin{equation}
A_{i} = \alpha(\xb_i)^2 A_0
\end{equation}
and $A_0$ is a constant. Here, $\alpha(\xb_i)$ is the predicted "action" by the RL algorithm.

In addition to the conservative force that arises from these potential terms, there is an active force
\begin{equation}
    F_{\rm active}(\xb) = (\alpha(\xb) \cos(\thetab), \alpha(\xb) \sin(\thetab), 0)
\end{equation}
where $\thetab$ is a vector of particle directions which itself has a purely diffusive dynamics
\begin{equation}
d\boldsymbol{\theta}_t = \sqrt{2 D_r} d\Wb_t.
\end{equation}

\begin{equation}
d\Xb_t = -[\nabla U_{\rm attract}(\Xb_t) - \nabla U_{\rm WCA}(\Xb_t) + F_{active}(\Xb_t, \boldsymbol{\theta}_t) ] dt +  \sqrt{2 D_t} d\Wb_t
\end{equation}

\begin{table}
\centering
\begin{tabular}{|p{3cm}||p{6cm}|}
 \hline
 \multicolumn{2}{|c|}{Simulation Parameters} \\
 \hline
 Parameter & Value (Normalized Units)\\
 \hline
 $D_{\rm r}$ & $0.125$\\
 $D_{\rm t}$ & $0.041667$\\
 $\epsilon$ & $0.5$\\
 $\sigma$ & $1$\\
 $A_0$ & $0.87$\\
 $\Delta t$ & $0.00005$\\
 \hline
\end{tabular}
\caption{Parameters for active colloid system.}
\end{table}

\section{Lennard-Jones System}\label{app:lj}

\begin{figure}
    \centering
    \includegraphics[width=0.45\linewidth]{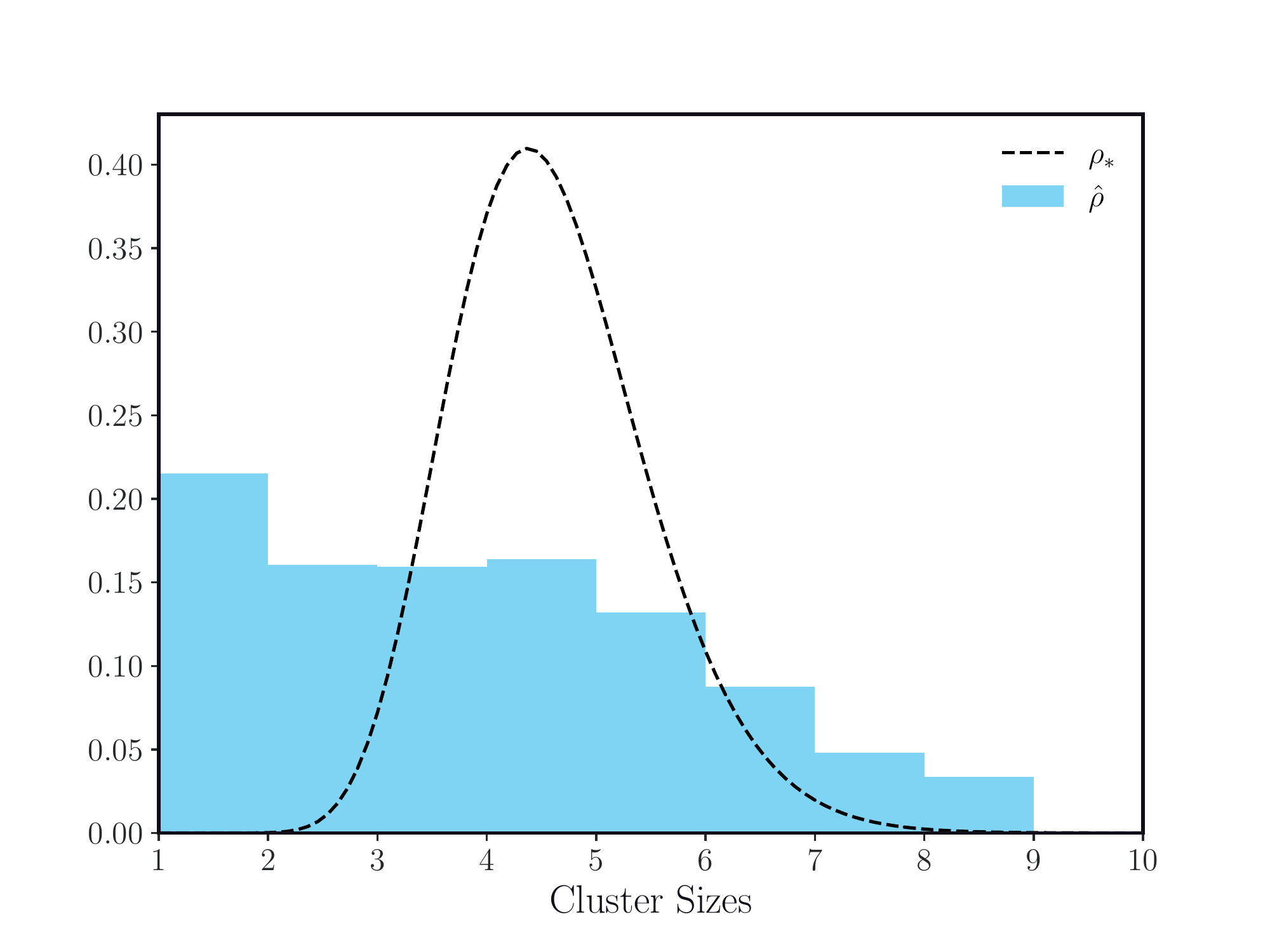}
    \includegraphics[width=0.45\linewidth]{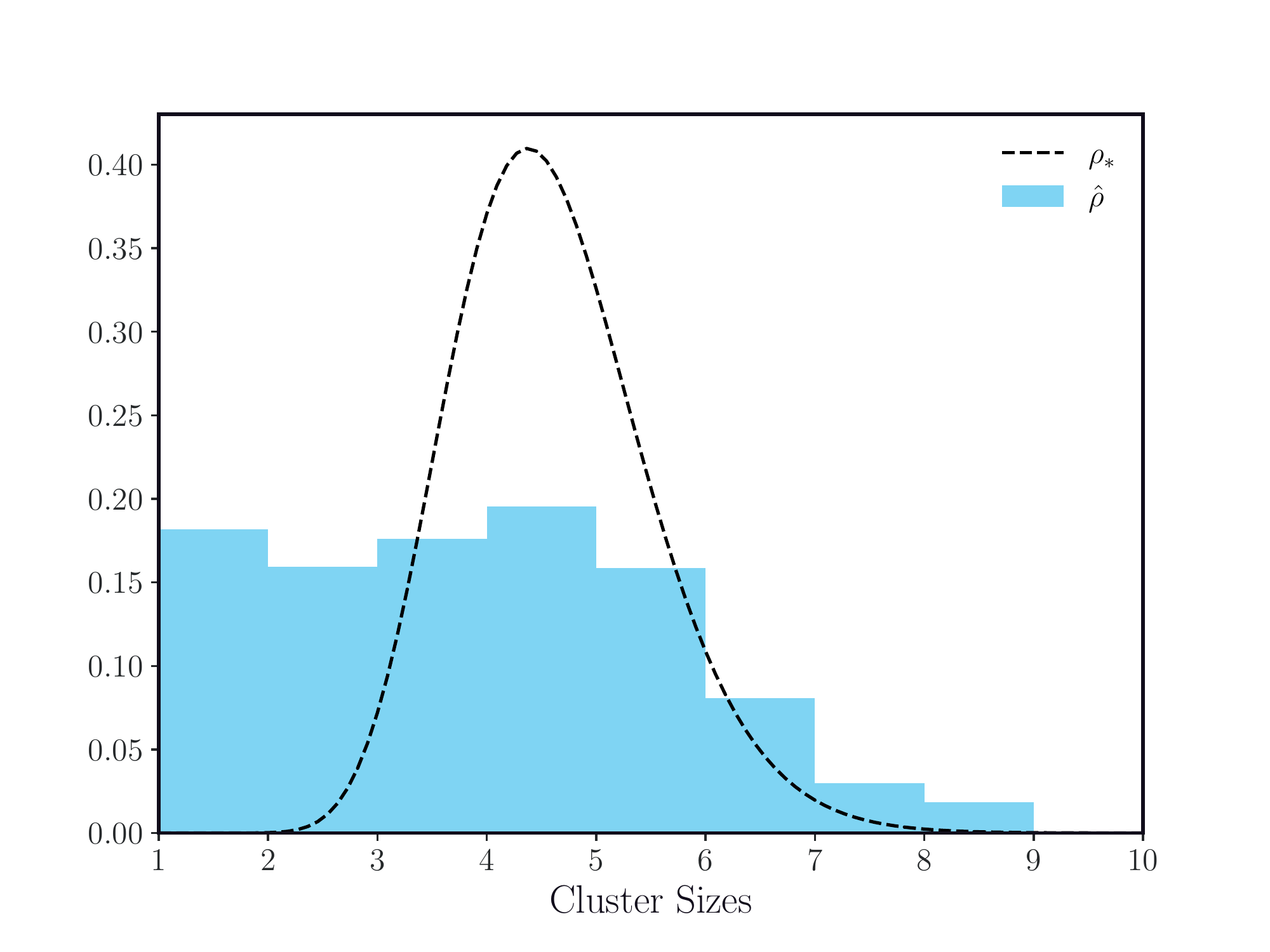}
    \includegraphics[width=0.45\linewidth]{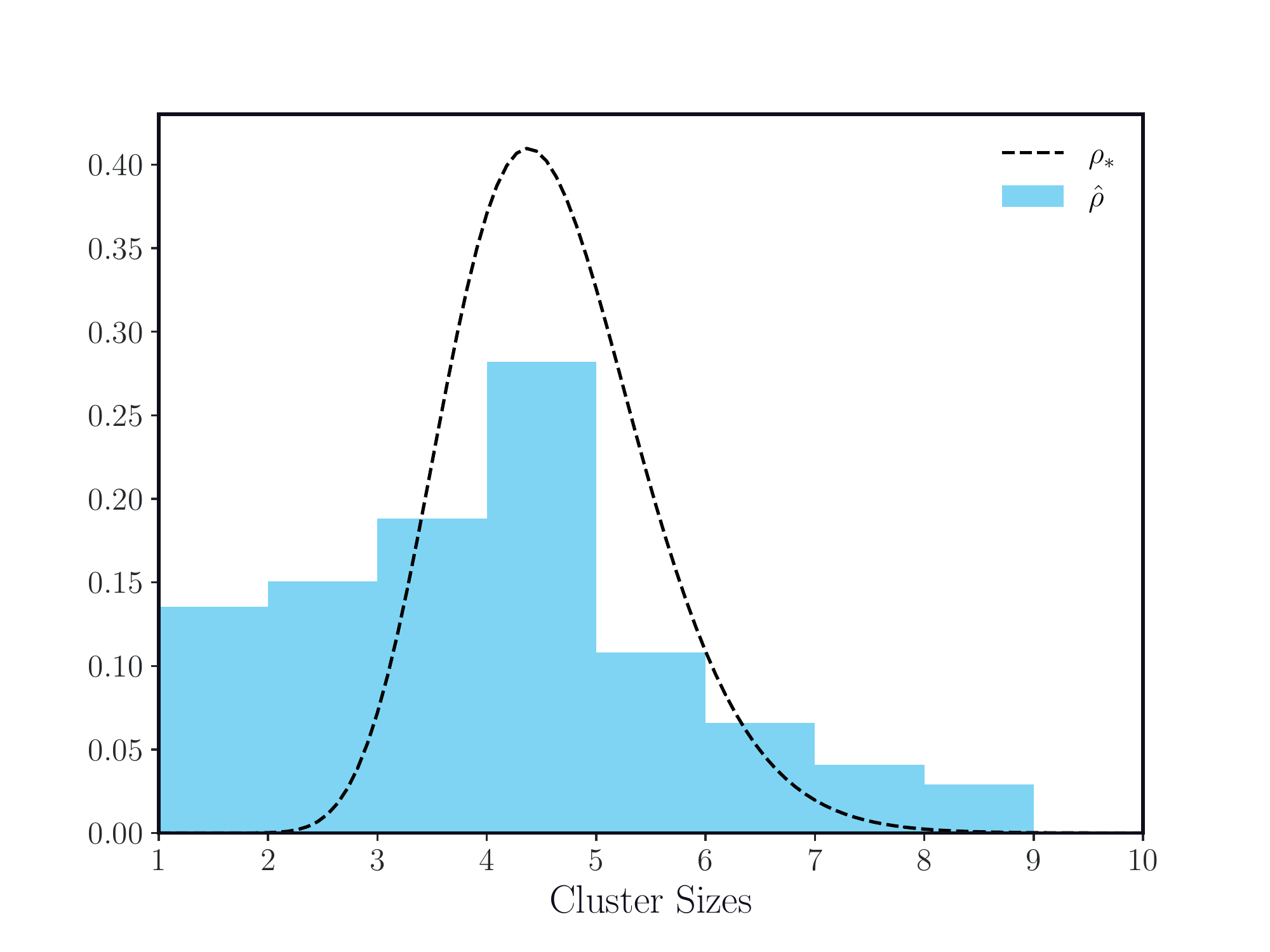}
    \includegraphics[width=0.45\linewidth]{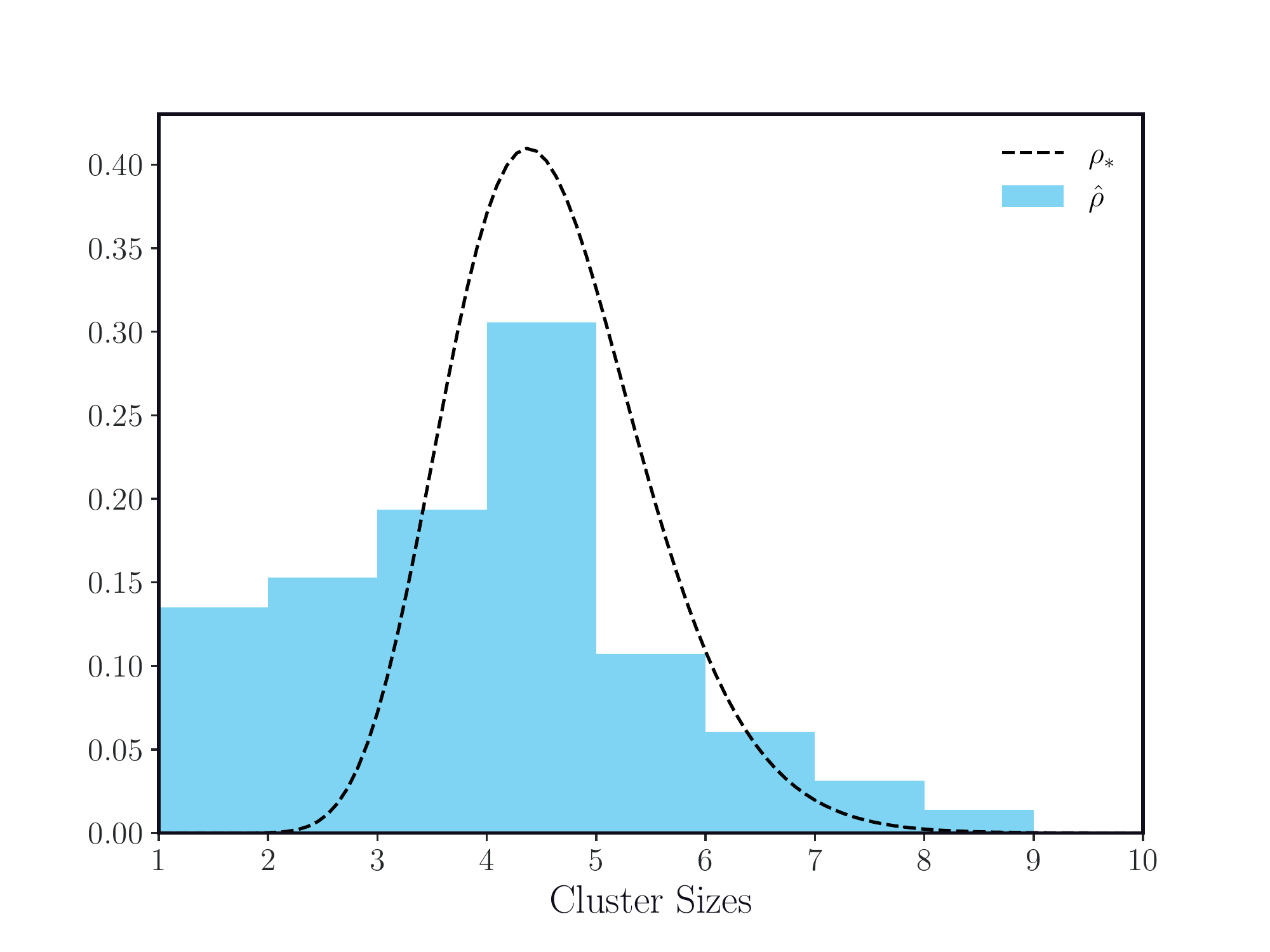}
    \caption{Histograms of cluster sizes for $3\times 3$ (top left) $4\times 4$ (top right) $12\times 12$ (bottom left) and $15 \times 15$ (bottom right).}
    \label{fig:ljsummary}
\end{figure}

To investigate feedback guided thermal annealing, we modeled a system of colloidal particles that interact via Lennard-Jones interactions.

\begin{equation}
U_{\rm LJ}(\xb_i, \xb_j) = 4 \epsilon \left[ \left(\frac{\sigma}{\| \xb_i - \xb_j \|}\right)^{12} - \left(\frac{\sigma}{\| \xb_i - \xb_j \|} \right)^{6} \right]
\end{equation}

Equation of Motion:
\begin{equation}
    d\Xb_t = -\nabla U_{\rm LJ}(\Xb_t) dt +  \sqrt{2 \beta^{-1}_t(\Xb_t)} d\Wb_t
\end{equation}

As described, we update the temperature of each grid of the system based on our Reinforcement Learning approach. The $\beta^{-1}_{t}$ for a particle depends on where the particle is at the beginning of a decision and is not changed during the duration of a decision. Because our decision length is only 0.25 seconds, our particle will not, on average, diffuse between grids within a decision even for our highest resolution of control. At the beginning of the next decision, we instantaneously update the temperature of each grid and subsequently update the $\beta^{-1}_{t + 1}$ for each particle depending on the temperature of the grid in which it is located.

\begin{table}[ht]
\centering
\begin{tabular}{|p{3cm}||p{6cm}|}
 \hline
 \multicolumn{2}{|c|}{Simulation Parameters} \\
 \hline
 Parameter & Value (Normalized Units)\\
 \hline
 $\epsilon$ & $0.5$\\
 $\sigma$ & $1$\\
 $\Delta t$ & $0.0002$\\
 \hline
\end{tabular}
\caption{Parameters for the thermal annealing simulations.}
\end{table}

\end{document}